\documentstyle[epsfig,amsfonts,amsbsy]{elsart}
\newcommand{\di}{{\mathrm d}}
\newcommand{\ii}{{\mathrm i}}

\newcommand{\for}{\quad{\mathrm{for}}\quad}
\newcommand{\with}{\quad{\mathrm{with}}\quad}
\newcommand{\where}{\quad{\mathrm{where}}\quad}
\renewcommand{\Re}{{\mathrm{Re}}}
\renewcommand{\Im}{{\mathrm{Im}}}
\def\lsim{\unitlength5mm\begin{picture}(1,0)
\put(0,.35){\makebox(1,0){$<$}}\put(0,0){\makebox(1,0){$\sim$}}
\end{picture}}
\def\eqs{\makebox(0,0){$=$}}
\def\pls{\makebox(0,0){$+$}}

\def\ssp{\makebox(0,0)
    {\thinlines\put(-.1,0){\line(1,0){.2}}\put(0,-.1){\line(0,0){.2}}}}
\def\ssm{\makebox(0,0){\put(-.1,0){\thinlines\line(1,0){.2}}}}
\def\photon{\thinlines\multiput(0,0)(.2,0){3}{\line(1,0){0.1}}}
\def\lphoton{\thinlines\multiput(0,0)(.2,0){5}{\line(1,0){0.1}}
\put(0.4,0){\vector(1,0){0.1}}}
\def\boson{\thinlines
           \multiput(0.0625,0)(.25,0){4}{\oval(.125,.125)[b]}
           \multiput(0.1875,0)(.25,0){4}{\oval(.125,.125)[t]}
           }

\def\Boson{\thicklines
           \multiput(0.0625,0)(.25,0){4}{\oval(.125,.125)[b]}
           \multiput(0.1875,0)(.25,0){4}{\oval(.125,.125)[t]}
           }
\def\VBoson{\thicklines
           \multiput(0.,-.625)(.0,.5){3}{\oval(.25,.25)[r]}
           \multiput(0,-.375)(.0,.5){3}{\oval(.25,.25)[l]}
           }
\def\vBoson{\thicklines
           \multiput(0.,-.3125)(.0,.25){4}{\oval(.125,.125)[r]}
           \multiput(0,-.4375)(.0,.25){4}{\oval(.125,.125)[l]}
           }
\def\vhBoson{\thicklines
           \multiput(-.5,-.375)(.25,.25){4}{\oval(.25,.25)[rb]}
           \multiput(-.25,-.375)(.25,.25){4}{\oval(.25,.25)[lt]}
           }
\def\hvBoson{\thicklines
           \multiput(.5,-.375)(-.25,.25){4}{\oval(.25,.25)[lb]}
           \multiput(.25,-.375)(-.25,.25){4}{\oval(.25,.25)[rt]}
           }

\def\Fermion{\thicklines\put(0,0){\vector(1,0){.6}}
            \put(0.6,0){\line(1,0){.4}}}
\def\fermion{\thinlines\put(0,0){\vector(1,0){.6}}
            \put(0.6,0){\line(1,0){.4}}}
\def\decomposition{
     \put(0,0){\photon}\put(0.3,0.3){\ssp}
     \put(2,0){\thicklines\oval(3.0,1.8)[l]
     \put(-.7,0){\makebox(0,0){$\beta^+$}}}
     \put(3,0){\thicklines\oval(3.0,1.8)[r]
     \put(.8,0){\makebox(0,0){$\alpha^-$}}}
     \put(4.5,0){\photon}\put(4.7,0.3){\ssm}
     \thicklines
     \put(2,-.9){\line(0,1){1.8}}
     \put(3,-.9){\line(0,1){1.8}}
     \put(2.5,0){\thicklines
     \multiput(0,0.25)(0,0.22){3}{\put(-.5,0){\vector(1,0){1}}
                            \put(-.7,0){\ssp}
                            \put(0.7,0){\ssm} }
     \multiput(0,-.25)(0,-.22){3}{\put(0.5,0){\vector(-1,0){1}}
                            \put(-.7,0){\ssp}
                            \put(0.7,0){\ssm} }}}
\def\oneloop{
     \put(1.5,0){\thicklines\oval(2.0,1.5)}
     \put(0,0){\photon}\put(0.3,0.3){\ssp}
     \put(2.5,0){\photon}\put(2.7,0.3){\ssm}}
\def\fullself{\begin{picture}(5,1)\put(0,0){\oneloop}
     \put(1.5,0){\makebox(0,0){$-\ii{\boldsymbol \Pi}$} }
     \end{picture}}

\def\oneloopvertex{
    \put(1.625,0){\thicklines\oval(2.0,1.5)}
    \put(0,0){\photon}\put(0.35,0.3){\ssp}
    \put(0.625,0){\circle*{.25}}\put(2.625,0){\circle*{.25}}
    \put(2.75,0){\photon}\put(2.9,0.3){\ssm}}
\def\Oneloopvertex{
    \put(1.625,0){\thicklines\oval(2.0,1.)}
    \put(0,0){\photon}\put(0.35,0.3){\ssp}
    \put(0.625,0){\circle*{.25}}\put(2.625,0){\circle*{.25}}
    \put(2.75,0){\photon}\put(2.9,0.3){\ssm}}
\def\mediumloop{
    \put(2.125,0){\thicklines\oval(3.0,1.5)}
    \put(0,0){\photon}\put(0.35,0.3){\ssp}
    \put(0.625,0){\circle*{.25}}\put(3.625,0){\circle*{.25}}
    \put(3.75,0){\photon}\put(3.9,0.3){\ssm}}
\def\longloop{
             \put(3.125,0){\thicklines\oval(5.0,1.5)}
             \put(0,0){\photon}\put(0.35,0.3){\ssp}
             \put(5.75,0){\photon}\put(5.9,0.3){\ssm}
             \put(0.625,0){\circle*{.25}}\put(5.625,0){\circle*{.25}}}
\def\doubleloop{
    \put(1.125,0){\thicklines\oval(1,2)}\put(2.625,0){\thicklines\oval(1,2)}
    \put(0,0){\photon}\put(0.35,0.3){\ssp}
    \put(0.625,0){\circle*{.25}}\put(3.125,0){\circle*{.25}}
    \put(3.25,0){\photon}\put(3.4,0.3){\ssm}
    \multiput(1.875,-.4)(0,.8){2}{\makebox(0,0){\rule{3mm}{1.5mm}}}
    \put(1.875,-.8){\ssm}\put(1.875,.8){\ssp} }
\def\Doubleloop{
    \put(1.125,0){\thicklines\oval(1,2)}\put(3.625,0){\thicklines\oval(1,2)}
    \put(0,0){\photon}\put(0.35,0.3){\ssp}
    \put(0.625,0){\circle*{.25}}\put(4.125,0){\circle*{.25}}
    \put(4.25,0){\photon}\put(4.4,0.3){\ssm}
    \multiput(1.875,-.4)(0,.8){2}{\makebox(0,0){\rule{3mm}{1.5mm}}}
    \multiput(2.875,-.4)(0,.8){2}{\makebox(0,0){\rule{3mm}{1.5mm}}}
    \multiput(2.375,-.4)(0,.8){2}{\thicklines\oval(.5,.7)}
    \put(1.875,-.8){\ssp}\put(1.875,.8){\ssp}
    \put(2.875,-.8){\ssm}\put(2.875,.8){\ssm} }

\def\borndiagram{\begin{picture}(2.5,1.5)
     \put(0,0){\borndiag}
     \put(.3,0){\ssm}\put(.3,1){\ssm}
     \put(1.225,0){\ssm}\put(1.225,1){\ssm}
     \put(2.7,0){\ssm}\put(2.7,1){\ssm}
     \put(2.,.5){\ssm}
     \end{picture}}
\def\borndiag{\begin{picture}(2.5,1.5)\thicklines
     \put(.25,.25){\vector(1,0){.75}}
     \put(1.125,.5){\makebox(0,0){\rule{2mm}{4mm}}}
     \put(1.,.25){\vector(1,0){1.75}}
     \put(.25,.75){\vector(1,0){.75}}\put(1.,.75){\vector(1,0){.9}}
     \put(2.,.75){\circle*{.2}}\put(2.,.75){\vector(1,0){.75}}
     \multiput(2.,.75)(0,.25){3}{\thinlines\line(0,1){.15}}\end{picture}}
\def\twocol{\begin{picture}(2.5,1.5)\thicklines
     \put(.25,0){\vector(1,0){.75}}
     \put(1.125,.375){\makebox(0,0){\rule{2mm}{6mm}}}
     \put(1.,0){\vector(1,0){.8}}
     \put(.25,.75){\vector(1,0){.75}}\put(1.,.75){\line(1,0){1}}
     \put(1.35,0){\borndiag}\end{picture}}
\def\selfinsert{\begin{picture}(5,2)\put(0,0){\Oneloopvertex}
     \put(1.2,.688){\fullbox}
     \put(1.2,.35){\ssp}\put(2.05,.35){\ssm}\put(2.05,.688){\fullbox}
     \put(.95,.875){\ssp}\put(2.3,.875){\ssm}
     \thicklines\put(1.625,.875){\oval(1.,.6)[t]}
     \put(1.2,.875){\line(1,0){.9}}
     \put(1.575,.5){\vector(1,0){.2}}
     \put(1.575,1.175){\vector(1,0){.2}}
     \put(1.675,-.5){\vector(-1,0){.2}}\put(1.675,.875){\vector(-1,0){.2}}
     \end{picture}}

\def\selfinsF{\begin{picture}(1.7,.6)\put(0.2,-.6){
     \put(.2,.688){\fullbox}\put(-.1,.5){\ssp}\put(1.4,.5){\ssm}
     \put(1.05,.688){\fullbox}
     \thicklines\put(.625,.875){\oval(1.,.6)[t]}
     \put(.2,.875){\line(1,0){.9}}
     \put(.1,.5){\line(1,0){1.1}}
     \put(.575,.5){\vector(1,0){.2}}
    \put(.575,1.175){\vector(1,0){.2}}
     \put(.675,.875){\vector(-1,0){.2}} }
     \end{picture}}
\def\boltzm{\begin{picture}(1.7,.6)\thicklines
\put(.2,-.1){\vector(1,0){1.3}}\put(.2,.275){\vector(1,0){1.3}}
\put(.85,.088){\fullbox}\end{picture} }
\def\nonskeleton{
     \thinlines\put(0,0){\line(0,1){1.75}}\put(0,0){\borndiagram}
          \put(2.,1.5){\thinlines\vector(0,1){.15}}
     \thinlines\put(3.,0){\line(0,1){1.75}}\put(3.25,1.65){\makebox(0,0){2}}
     \put(4.75,.8){\makebox(0,0){$\longrightarrow$}}
     \put(6.25,.8){\selfinsert}}
\def\fullvertexeq{\begin{picture}(8.5,1)
    \put(.2,0){
    \put(0,0){\lphoton}
    \put(0.9,0){\thicklines\vector(1,1){.625}}
    \put(1.525,-.625){\thicklines\vector(-1,1){.55}}
    \put(.9,0){\circle*{.25}}}\put(.7,.3){\ssm}
    \put(2.5,0){\makebox(0,0){$=$}}
    \put(3,0){
       \put(0,0){\lphoton}
       \put(.7,.3){\ssm}\put(.9,0){\thicklines\vector(1,1){.7}}
       \put(1.6,-.7){\thicklines\vector(-1,1){.7}}}
       \put(5.5,0){\makebox(0,0){$+$}}
    \put(6,0){\put(0,0){\lphoton}\put(.7,.3){\ssm}
      \put(1.5,0){\thicklines\oval(1.25,.5)[l]}
      \put(1.2,.25){\thicklines\vector(1,0){.2}}
      \put(1.6,.25){\thicklines\vector(1,1){.5}}
      \put(2.1,-.75){\thicklines\vector(-1,1){.5}}
      \put(1.475,0){\thicklines\makebox(0,0){\rule{.25cm}{.5cm}} }
      \put(1.475,.55){\ssm}\put(1.475,-.55){\ssm}}
   \end{picture}}

\def\Fullbox{\thicklines\put(0.5,0){\vector(-1,0){.5}}
                \put(1.5,0){\vector(-1,0){.5}}
                \put(0,1){\vector(1,0){.5}}\put(.5,1){\vector(1,0){1}}
                \put(.75,.5){\makebox(0,0){\rule{.5cm}{1.cm}}}
                \put(.75,1.2){\ssm}\put(.75,-.2){\ssm}}
\long\def\fourpointeq{\begin{picture}(14,1.2)
     \put(0,.25){\Fullbox}\put(2,.75){\eqs}
     \put(2.5,.25){\thicklines\put(.625,0){\vector(-1,0){.625}}
              \put(0,1){\vector(1,0){.625}}
              \put(1.5,0){\vector(-1,0){.625}}
              \put(.875,1){\vector(1,0){.625}}
              \put(.75,.5){\vBoson}
              \put(.75,0){\circle*{.2}}\put(.75,1){\circle*{.2}}
              \put(.75,1.2){\ssm}\put(.75,-.2){\ssm}}
     \put(4.5,.75){\pls}
     \put(5,.25){\thicklines\put(1.625,0){\vector(-1,0){1.625}}
           \put(0,1){\vector(1,0){.625}}
           \put(2.5,0){\vector(-1,0){0.625}}
           \put(.875,1){\vector(1,0){1.625}}
           \put(.75,1.2){\ssm}\put(.75,-.2){\ssm}
           \put(1.75,1.2){\ssm}\put(1.75,-.2){\ssm}
           \put(.75,.5){\vBoson}\put(1.75,.5){\vBoson}
           \put(.75,0){\circle*{.2}}\put(.75,1){\circle*{.2}}
           \put(1.75,0){\circle*{.2}}\put(1.75,1){\circle*{.2}}  }
     \put(8,.75){\pls}
     \put(8.5,.25){\thicklines\put(1.625,0){\vector(-1,0){1.625}}
           \put(0,1){\vector(1,0){.625}}
           \put(2.5,0){\vector(-1,0){0.625}}
           \put(.875,1){\vector(1,0){1.625}}
           \put(1.25,.5){\vhBoson}\put(1.25,.5){\hvBoson}
           \put(.75,0){\circle*{.2}}\put(.75,1){\circle*{.2}}
           \put(1.75,0){\circle*{.2}}\put(1.75,1){\circle*{.2}}
           \put(.75,1.2){\ssm}\put(.75,-.2){\ssm}
           \put(1.75,1.2){\ssm}\put(1.75,-.2){\ssm}}
     \put(12,.75){\makebox(0,0){$\dots$}}\end{picture}}
\def\twodecomp{\begin{picture}(8,2)
     \put(0,1.1){
      \put(0,0.){\oneloop}\put(1.25,0){\VBoson}\put(1.75,0){\VBoson}
      \put(1.25,1.){\ssp}\put(1.25,-1.){\ssm}
      \put(1.75,1.){\ssm}\put(1.75,-1.){\ssp}
      \thinlines\put(0.5,-.75){\line(2,3){1.08}} }
     \put(4,1.1){
      \put(0,0){\oneloop}\put(1.25,0){\VBoson}\put(1.75,0){\VBoson}
      \put(1.25,1.){\ssp}\put(1.25,-1.){\ssm}
      \put(1.75,1.){\ssm}\put(1.75,-1.){\ssp}
      \thinlines\put(2.5,-.75){\line(-2,3){1.08}}}
      \end{picture}}
\def\DysonF{\begin{picture}(8,0)
\put(0,0){\Fermion}\put(1.5,0){\eqs}\put(2,0){\fermion}\put(3.5,0){\pls}
\put(4,0){\fermion}\put(5.75,0){\thicklines\oval(1.5,1.0)}
\put(5.75,0){\makebox(0,0){$-i{\boldsymbol \Sigma}_F$}}\put(6.5,0){\Fermion}
        \end{picture} }

\def\DysonB{\begin{picture}(8,0)
    \put(0,0){\Boson}\put(1.5,0){\eqs}\put(2,0){\boson}\put(3.5,0){\pls}
    \put(4,0){\boson}\put(5.75,0){\thicklines\oval(1.5,1.0)}
   \put(5.75,0){\makebox(0,0){$-i{\boldsymbol
 \Sigma}_B$}}\put(6.5,0){\Boson}
   \end{picture}}

\def\fullbox{\makebox(0,0){\rule{1.5mm}{3mm}}}
\def\interaction{\makebox(0,0){\put(0,0){\interact}
    \put(0,.95){\ssp}\put(0,-.95){\ssm}
    \put(0,.125){\ssp}\put(0,-.125){\ssm}}}
\def\interact{\makebox(0,0){\put(0,.5){\fullbox}
    \thicklines\put(0,0){\oval(.75,.5)}
    \put(0,-.5){\fullbox}} }
\def\intrad{\begin{picture}(3,1.5)\put(1,.75){\interact}\thicklines
     \put(0,0){\vector(1,0){.9}}\put(0,1.5){\vector(1,0){.9}}
     \put(.9,0){\vector(1,0){1.1}}\put(.9,1.5){\vector(1,0){1.1}}
     \put(1.375,.75){\lphoton}\end{picture} }
\def\til2loop{\put(0,0){\oneloopvertex}\put(4.,0){\pls}
      \put(4.75,0){\oneloopvertex}\put(6.375,0){\interaction}
      \put(9,0){\pls}}
\def\classdiagram{
     \begin{picture}(18,.8)
     \put(0,0){\til2loop}\put(9,0){\nloop}
     \put(16,0){\pls}\put(17,0){\makebox(0,0){$\dots$}}
     \end{picture}}
\def\keydiagrams{\begin{picture}(21,4)\put(0,3){
   \put(0,0){\fullself}\put(3.5,0){\eqs}\put(4,0){\til2loop}}
   \put(13.75,3){\put(0,0){\mediumloop}
      \multiput(1.625,0)(1,0){2}{\interaction}}
      \put(18.75,3){\pls}\put(19.75,3){\makebox(0,0){$\dots$}}
   \put(-1,0){
   \put(4,.2){\doubleloop}\put(8.5,.2){\pls}\put(3.25,.2){\pls}
   \put(9.25,.2){\put(0,0){\mediumloop}
      \thicklines\multiput(1.615,-.75)(.02,0){10}{\line(2,3){1}}
      \multiput(2.615,-.75)(.02,0){10}{\line(-2,3){1}}
      \put(1.625,.95){\ssp}\put(1.625,-.95){\ssm}
      \put(2.625,.95){\ssm}\put(2.625,-.95){\ssp}}
   \put(14.25,.2){\pls}\put(15,.2){\Doubleloop}
   \put(20.5,.2){\pls}\put(21.5,.2){\makebox(0,0){$\dots$}}}
   \end{picture}}
\def\nloop{\begin{picture}(6.5,.5) \put(0,0){\longloop}
       \multiput(1.625,0)(1,0){2}{\interaction}\put(3.625,0)
       {\makebox(0,0){$\dots$}}\put(4.625,0){\interaction}\end{picture}}

\makeatletter
\def\ps@copyright{\let\@mkboth\@gobbletwo
  \def\@oddhead{}%
  \let\@evenhead\@oddhead
  \def\@oddfoot{\small\sl
      GSI-Preprint 95-63, subm. to Ann. Phys.,
      (hep-ph/95****)\hfill }
  \let\@evenfoot\@oddfoot
}
\begin{document}
\begin{frontmatter}
\title{CLASSICAL AND QUANTUM MANY-BODY DESCRIPTION\\
OF BREMSSTRAHLUNG IN DENSE MATTER\\
(Landau - Pomeranchuk - Migdal Effect)}
\author{J\"orn Knoll and Dmitri N. Voskresensky\thanksref{prmai}}
\address{
        Gesellschaft f\"ur Schwerionenforschung GSI\\
        P.O.Box 110552, D-64220 Darmstadt, Germany}
\thanks[prmai]{permanent address: Moscow Institute for Physics and
Engineering, Russia, 115409 Moscow, Kashirskoe shosse 31\\
Electronic mail: J.Knoll@gsi.de, voskre@rzri6f.gsi.de}
\begin{abstract}
Some considerations about the importance of coherence effects for
bremsstrahlung processes in non--equilibrium dense matter (Landau -
Pomeranchuk - Migdal - effect) are presented. They are of particular
relevance for the application to photon - and di-lepton production
from high energy nuclear collisions, to gluon radiation in QCD
transport, or parton kinetics and to neutrino and axion radiation from
supernova explosion and from hot neutron stars. The soft behavior of
the bremsstrahlung from a source described by classical transport
models is discussed and pocket correction formulas for the in-matter
radiation cross sections are suggested in terms of standard transport
coefficients. The radiation rates are also discussed within a
non--equilibrium quantum field theory (Schwinger - Kadanoff - Baym -
Keldysh) formulation. A classification of diagrams and corresponding
resummation in physically meaningful terms is proposed, which
considers the finite damping width of all source particles in matter.
This way each diagram in this expansion is already free from the
infra--red divergences.  Both, the correct quasi--particle and
quasi--classical limits are recovered from this subset of
graphs. Explicit results are given for dense matter in thermal
equilibrium. The diagrammatic description may suggest a formulation of
a transport theory that includes the propagation of off--shell
particles in non--equilibrium dense matter.
\end{abstract}
\end{frontmatter}
\clearpage

\section{Introduction}

The importance of coherence time effects on the production and
absorption of field quanta from the motion of source particles in
non-equilibrium dense matter has first been discussed by Landau,
Pomeranchuk, Migdal (LPM) \cite{LandauP,Migdal} (and many others
later) in the context of bremsstrahlung from ultra--relativistic
electrons undergoing multiple rescatterings on Coulomb centers. The
first successful measurements of the corresponding suppression of
bremsstrahlung have been carried out at the Stanford Linear
Accelerator Center very recently \cite{eSLAC}. With this
paper\footnote{a brief report of these results is given in \cite{KV}}
we like to supplement some quite intuitive and also formal
considerations, which illustrate the nature of production and
absorption processes in a dense matter environment.  The subject is of
quite general nature and applies to many physical problems, where
either a source couples weakly to a wave field or for the proper
determination of local gain and loss terms in quantum transport.
Examples are the application to photon, or di-lepton production from a
piece of dense nuclear matter or hadron gas formed in high energy
nuclear collisions, for gluon or parton radiation and absorption in
QCD transport and its practical implementation in parton kinetic
models (such problems are discussed, e.g. in \cite{WangGy,Proc}), to
neutrino and axion radiation from supernovas and neutron-star matter
(see \cite{ST,NS,RS,MSTV}), for the soft phenomena in quantum
cosmological gravity (see \cite{TW}) and also for many condensed
matter phenomena, as particle transport in metals and semiconductors,
radiation in plasma etc. (see \cite{SL,RS}). To be specific, however,
we take the example of electrodynamics, considering photon production
from a piece of nuclear matter, but when appropriate comment on other
cases. Since throughout the paper we discuss the corresponding {\em
proper self energy} of the produced particle, all considerations given
also apply to gain and loss terms of other particles in
non-equilibrium dynamics.

In the context of high-energy nucleus--nucleus collisions \cite{Proc}
for example, it became quite apparent over the last years, that a
justification of QCD transport (e.g. in terms of a parton kinetic
picture) calls for a proper understanding of all soft processes. Well
known is the Rutherford singularity in scattering cross-sections of
interactions mediated by the exchange of zero mass quanta (photons/
gluons). In dense matter the exchanged quantum acquires a finite real
mass due to Debey-screening.  Singularities are also encountered in
absorption or radiation processes (bremsstrahlung). Induced by free
scattering the rates diverge at vanishing four-momentum $q$ of the
radiated quantum, due to the infinite time scales used in the
quasi-free approximation. In dense matter, however, due to the {\em
finite free propagation time} $\tau_{\rm coll}$ between successive
collisions these rates become regular.  In particle physics context,
most of the papers on the LPM effect discuss the brems\-strahlung of
some fast charged particle, such as hadron, quark or gluon, which
traverses a dense hadron gas or quark--gluon plasma, e.g. see refs.
\cite{Cleymans,WangGy}, where the role of the matter is reduced to
infinitely massive scatterers. In reality all the particles in dense
matter which couple to the radiated field should be treated on equal
footing. Effects of the finite mean free propagation time on photon
and gluon radiation have been considered e.g. in refs.
\cite{Cleymans,DGK,KnollLenk,BDRS,WangGy}.

While the problem can be quite simply and intuitively formulated and
solved in the classical limit, where a classical source couples to a
wave field, e.g.  classical charge particles couple to a Maxwell
field, considerable conceptual difficulties arise for the very same
problem, if the source is described as a quantum many-body system. In
fact common standard techniques, like perturbation theory or the
quasi-particle approximation (QPA) have serious limitations to
describe the production and/or absorption rates over the whole range
of energies and momenta, as they completely fail in the soft limit.

Starting from a quantum many-body formulation in terms of Green's
functions most derivations of transport descriptions employ two
essential approximation steps: i) a gradient expansion and ii) the
QPA. For simplicity we concentrate on the defects of the QPA in this
paper. In the QPA, which is a consistent approximation scheme for low
temperature Fermi liquids (Landau - Migdal, see \cite{LL,M}), all
particles in the medium are treated {\em on-shell} with a well
determined energy-momentum relation (dispersion relation) which
follows from the real part of the retarded self energy of the
particle. To be specific in this notion, we use the term ''on-shell'',
when the particle follows a sharp energy-momentum relation. Thus the
quasi-particle poles of the retarded Green's function lie just
infinitesimally below the causality cut along the real axis in
energy. The corresponding approximation scheme in terms of these
on-shell states, which have infinite life time, is called
quasi-particle approximation (QPA).  Due to interactions in dense
matter the {\em damping} of the quasi-particles may become important,
the corresponding ''quasi-particle'' poles of the retarded propagators
move into the unphysical sheet below the real axis. As a consequence
the mass spectrum of the particles is no longer a sharp delta function
but rather acquires a width $\Gamma$, and one talks about
''off-shell'' propagation. In that case one has to leave the standard
description in terms of stable single particle states and employ
quantum propagators (Green's functions) with continuous mass
distributions. Landsmann \cite{Landsmann} has coined the notion of
''non-shell particles'' in this connection.  One thus comes to a
picture which unifies {\em resonances} which have already a width in
vacuum due to decay modes with the ''states'' of particles in dense
matter, which obtain a width due to collisions (collisional
broadening).

The theoretical concepts for a proper many body description in terms
of a real time non equilibrium field theory have already been devised
by Schwinger, Kadanoff, Baym and Keldysh \cite{Schwinger,KB,Keldysh}
in the early $60^{ies}$. First investigations of the quantum effects
on the Boltzmann collision term were given by Danielewicz \cite{D},
the principle conceptual problems on the level of quantum field theory
were investigated by Landsmann \cite{Landsmann}, while applications
which seriously include the finite width of the particles in transport
descriptions were carried out only in recent times,
e.g. \cite{SL,D,BDG,BM,SCFNW,VBRS,PHenning1,PHenning,DB,HFN,HS,QH}.
For resonances, e.g. the delta resonance in nuclear matter, it was
natural to consider broad mass distributions and ad hoc recipes have
been invented to include this in transport simulation models. However,
many of these recipes are not correct as they violate some basic
principle like detailed balance \cite {DB}, and the description of
resonances in dense matter has to be improved. The present study also
gives some hints on how to generalize the transport picture towards
the inclusion of off-shell propagations in dense matter
\cite{BM,PHenning1,PHenning}.

In this paper we illustrate the practical implications of such
non-equilibrium concepts at the example of particle production from
the dense matter dynamics. Thus all source particles never reach an
asymptotic state and naturally have a continuous mass spectrum. In
sect. 2 we derive the basic formulas for the rate of
bremsstrahlung. The classical and general quantum mechanical
expressions, the latter in terms of non--equilibrium Green's functions
and self energies, are derived for the case of non--equilibrium
dynamics. In sect. 3 we concentrate on the description of radiation
from classical sources. We start with the bremsstrahlung from a
classical diffusion process, and subsequently derive the photon
spectrum for a classical random walk (Langevin) process in terms of a
completely regular multiple collision expansion.  The low energy
behavior is discussed and pocket correction formulas for the
in--matter radiation cross sections are suggested in terms of standard
transport coefficients. Also finite size corrections are
obtained. Then in sect. 4 we use the non--equilibrium Green's function
formalism, see \cite{Schwinger,KB,Keldysh,D,LP,VBRS}, and formulate
diagrammatic resummations where all quantities are expressed through
physically meaningful terms. We show how infra-red convergent results
can be obtained through the account of the finite damping width and
discuss the QPA (sect. 4.5) and quasi--classical (QC) (sect. 5) limits
from the corresponding infinite series of diagrams.  In sect. 6 the
lowest order loop diagrams for the production rate from a piece of
equilibrium dense matter are analyzed in the quantum case, both at
high and low temperatures. Conclusions and perspectives are given in
sect. 7. Some formal details are deferred to the Appendix.

We use rational units $\hbar=c=1$. Whenever the behavior of some
quantity is discussed in the classical limit ($\hbar\rightarrow 0$),
$\hbar$ will be given explicitly.

\section{Basic Formulas for the Rate of Bremsstrahlung}

If the source system couples only perturbatively (to lowest order in
$e^2$) to the electromagnetic field, the production or absorption rate
of photons can be formulated using standard text book concepts in
terms of Fermi's golden rule. The corresponding transition amplitude
is given by the electromagnetic current operator between the initial
and final states of the source. For dense matter problems it is more
advantage to use a more general concept, where the local production
and absorption rates are expressed through the current-current
correlation function\footnote{For the description of coordinates and
momenta we use the following conventions: numbers $1$, $2$,
etc. abbreviate space-time points $x_1=(t_1,{\vec x_1})$, etc.; for
two-point functions coordinate means are $x=(x_1+x_2)/2=(t,{\vec x})$,
relative coordinates: $\xi=x_1-x_2=(\tau,{\vec \xi})$; the
corresponding four vector Wigner coordinates are $(x;q)=(t,{\vec
x};\omega,{\vec q})$ for the photon and $(x;k)=(t,{\vec
x};\epsilon,{\vec k})$ for the particles of the source. Whenever
advantage or necessary we shall swap from one to another or even to
some mixed representation, just changing the corresponding arguments
of the functions; for space -, or space-time independent systems we
drop the argument ${\vec x}$ or $x$, respectively; e.g.:
$\Pi_{12}=\Pi(1;2)\leftrightarrow \Pi(x;q)\rightarrow \Pi(\omega,{\vec
q})\leftrightarrow \Pi(\tau ,{\vec q})$; the latter two in space-time
homogeneous systems.  For simplicity the polarization indices $\mu$
and $\nu$ or $i$ and $k$ for the spatial part of the tensor structure
of $\Pi^{\mu\nu}$ will not always be given in later equations.}
\begin{eqnarray}
\left<j^{\nu\dagger}(2)j^{\mu}(1)\right>
\quad\mbox{for production, and}\quad
\left<j^{\nu}(2)j^{\mu\dagger}(1)\right>
\quad\mbox{for absorption}.
\end{eqnarray}
Although $j(x)=j({\vec x},t)$ is a hermitian operator we distinguish
between $j$ and $j^{\dagger}$ in order to designate the photon
creation and annihilation vertex, respectively.  The bracket
$\left<\dots \right>$ denotes a quantum ensemble average over the
source; quantum states and operators are taken in the Heisenberg
picture.

With reference to the description of non--equilibrium systems, where
it is advantage to use real-time non--equilibrium field theory
concepts, such as the Schwinger - Kadanoff - Baym - Keldysh technique
\cite{Schwinger,KB,Keldysh,D} we introduce the following notions
\begin{eqnarray} \label{Npi-+}
4\pi\left<j^{\nu\dagger}(2)
j^\mu(1)\right> = -\ii\Pi^{\mu\nu -+}(1;2),\quad
4\pi\left<j^{\nu}(2)
j^{\mu\dagger}(1)\right> = -\ii\Pi^{\nu\mu +-}(2;1)\; ,
\end{eqnarray}
which relate the correlation functions to the proper self energies
$\Pi^{-+}$ and $\Pi^{+-}$ of the photon, which are responsible for
gain and loss (c.f.  sect. 4.3). Throughout this paper we use the
$\{-,+\}$ notation, defined in detail in sect. 4 in the convention
of ref.  \cite{LP}, chapt. 10.

In this formulation the production term for the phase space occupation
$n_{\gamma}({\vec x},{\vec q},t)$ (Wigner density) of on-shell photons
per space--time volume $\di^4 x=\di^3x\;\di t$, and per
energy--momentum $\omega -{\vec q}$ volume, $\di^4q=\di\omega\di^3 q$,
with polarization $ \eta=\{\eta_\mu\}$ is given by

\begin{equation} \label{dN}
   \di^8 n_{\gamma}({\vec x},{\vec q},t) =
   -\ii\eta_{\mu}\eta_{\nu}\Pi^{\mu\nu-+}(x;q)\,
 (1+n_{\gamma}({\vec x},{\vec q},t))\,
 \delta(\omega^2 -\omega_{\vec q}^2)\di^4x \di^4 q\, ,
\end{equation}
where $\omega_{\vec q}$ is the photon on-shell energy, and
\begin{eqnarray} \label{Pi-+xq}
-\ii\Pi^{\mu\nu-+}(x;q)&=& 4\pi\int \di^4 \xi
  e^{\ii q\xi}\left<j^{\nu\dagger}(x-\xi/2)
   j^\mu(x+\xi/2)\right>
\end{eqnarray}
denotes the space-time Wigner transformation of the auto correlation
function (\ref{Npi-+}).  This local gain term is the on-shell version
of a general quantum transport concept (Kadanoff-Baym equation
\cite{KB}, c.f. (\ref{KBeq}), sect. 4). It likewise applies for
virtual photons (e.g. dilepton production), replacing the on-shell
$\delta$-function in (\ref{dN}) by the corresponding off-shell photon
spectral function.

The above expressions are the space and time-dependent version of the
more familiar golden rule for quantum transitions between exact
stationary eigenstates. Note that by definition, c.f. (\ref{Pi-+xq}),
$-\ii\eta_{\mu}\eta_{\nu}\Pi^{\mu\nu}$ is a real quantity; if integrated
over phase-space volumes $\Delta x\Delta q$ large compared to $\hbar$
it becomes positive and serves as a production rate.

Such a formalism has been applied in many cases employing the QPA for
the equilibrium Green's functions, c.f. refs. \cite{V1,MSTV}. However,
the general formalism allows to go beyond this limit and to account
for the finite damping width of the source particles due to their
finite mean free path, which is the main topic of this paper.

Therefore the current-current correlation function is the central
quantity of interest. In graphical form it is determined by the proper
self energy diagram of the photon
\begin{equation}\label{grafpi}\unitlength6mm
   -\ii\Pi^{-+}=
   \begin{picture}(5,.7)
   \put(0.2,0.2){\fullself}
   \end{picture}
\end{equation}
which sums all one-photon irreducible self energy
diagrams\footnote{To order $e^2$ naturally all diagrams are one-photon
irreducible; however for the application to the production of
particles with a larger coupling constant, e.g. for gluons with
coupling constant $g$, also diagrams to higher order in $g$ are
relevant and one then has to discard diagrams which are one gluon line
reducible.}.  The dashed lines relate to the photon, while the
interior area ($-\ii\Pi$) symbolically denotes the exact inclusion of
all strong interactions among the source particles.

\subsection{Analytical properties and constraints}

The self energies for gain and loss obey some analytical relations
that follow right from the definitions (\ref{Npi-+},\ref{Pi-+xq}), like
$$-\ii\Pi^{\mu\nu -+}(1;2)=\left(-\ii\Pi^{\nu\mu -+}(2;1)\right)^*$$
which implies that $-\ii\eta_{\mu}\eta_{\nu}\Pi^{\mu\nu}(x;q)$ is
real.  Production and absorption parts obey
$$\Pi^{\mu\nu -+}(x;q)=\Pi^{\mu\nu +-}(x;-q).$$ Integration over
$\omega$ projects onto equal time properties. Of particular help for
the discussion of soft processes \cite{DGK,KnollLenk} are the
following energy weighted dipole (${\vec q}=0$) sum-rules
(e.g. \cite{BetheS})
\begin{equation}\label{sumrule}
   \hspace*{-1cm}
   -\ii\int_{-\infty}^{\infty}
 \frac{\di\omega}{2\pi}\omega^{n-2}\Pi^{-+}(\omega,{\vec
   q}=0;t,{\vec x})\di^3x=S_n=\left\{\begin{array}[c]{ll}
   4\pi\left<J^i(t)J^k(t)\right> &{\for}n=2\\
   -2\pi \ii\left< [D^i(t),J^k(t)]\right> &{\for}n=1\quad\\
   4\pi\left<D^i(t)D^k(t)\right> &{\for}n=0\end{array}\right.
\end{equation}
which are valid also in the general non-equilibrium case. Here
$i,k\in\{1,2,3\}$ denote the spatial components. The r.h.  expressions
are given by the space integrated currents and dipole moments
\begin{equation}
  J^i(t)=\int\di^3 x j^i(t,{\vec x});\quad\quad
  D^i(t)=\int\di^3 x x^ij^0(t,{\vec x}).
\end{equation}
While the $n=2$ sum-rule directly follows from definition
(\ref{Pi-+xq}) and applies to {\em any} current even non conserving,
the other two use current conservation and partial integrations, known
as Siegert's theorem (c.f. \cite{DeshalitF}), and therefore also
require that the system has a finite space extension.  (If there are
no long range correlations, the $n=2$ and $n=1$ relations can also be
used for infinite matter, if properly taken per volume). For
non-relativistic currents the commutator in the $n=1$ (Thomas - Reiche
- Kuhn) sum-rule just becomes the sum of square charges in the system
$\ii [D^i(t),J^k(t)]=\sum e^2_{\nu}$, where $\nu$ labels the
constituents.

For systems in {\em thermal equilibrium} production and absorption follow
the detailed balance relation (Kubo - Martin - Schwinger \cite{KMS})
\begin{equation}\label{KMS}
\Pi^{-+}(q;x)=\Pi^{+-}(q;x)e^{-\omega/T}
\end{equation}
where $T$ is the temperature.  They allow to write the
l.h.s. of the sum-rules as half-sided integrals, e.g. integrating
only the production rate
\begin{equation}
  -\ii\int_0^{\infty}\frac{\di\omega}{2\pi}\omega^{n-2}
   \Pi^{-+}(\omega,{\vec q}=0;t,{\vec x})
   \left(1+(-1)^ne^{\omega/T}\right)\di^3x=S_n.
\end{equation}
These rules have been used to estimate the validity of the quasi-free
scattering prescription in kinetic models \cite{DGK,KnollLenk}.  In
the classical limit, where $\hbar\omega\rightarrow 0$
(c.f. sect. 3. below) the l.h.s. of the $n=1$ and $n=2$ sum-rules
coincide in equilibrium, apart from a factor $T/2$, and the ensuing
identity $2 S_2=T S_1$ is a disguised form of the classical equal
partition theorem.

A further consistency check for diagrammatic elements which determine
the self energy can be given in terms of Ward identities in the case
of conserved currents. Since the space-integral of $j^0$ gives the
conserved total charge $Z$ of the system, one may also use that the
space integrated density-density correlator is constant in time, i.e.
\begin{equation}\label{timewardid}
 \int\di^3 x_1\di^3 x_2\Pi^{00 -+}(1;2)=\int\di^3 x\Pi^{0 \nu
-+}(\tau,{\vec q}=0;t,{\vec x})= 4\pi Z^2=const.,
\end{equation}
which applies even in non-equilibrium cases. For isolated systems the
motion of the center of mass leads to no radiation. Therefore one
normally introduces effective charges for the different kind of
particles of the source in the standard manner such that the total
effective charge vanishes $Z_{eff}=0$.

\section{Radiation from Classical Sources}\label{sect-class}

In this section we discuss two examples which treat the source as a
classical system coupled to a Maxwell field. This limit just amounts
to evaluate the current-current correlator on the classical
level\footnote{One has to realize that a classical photon carries no
energy in the quantum sense, i.e. $\hbar\omega\rightarrow 0$ and the
energy is given by the electromagnetic fields.}.  We discuss the
radiation caused by a single charged particle (the source), which
stochastically moves in neutral dense matter. The motion of the source
is described ($a$) by mesoscopic transport (diffusion process) and
($b$) by a microscopic Langevin process. Since these examples
represent the QC limits of the corresponding quantum field theory
cases, we carry on the discussion in terms of the photon self energy
$\Pi^{- +}$.

\subsection{Diffusion Process}

The motion of a non-relativistic source particle is assumed to be
described by a time dependent phase-space distribution $f({\vec
x},{\vec v},t)$ in space and velocity with convective current density
${\vec j}({\vec x},t)= e\int \di^3 v\;{\vec v}\;f({\vec x},{\vec
v},t)$.  For standard dissipative media in equilibrium the velocity
autocorrelation function (integrated over space) decays exponentially
in time

\begin{equation}\label{vrelax} \left< v^i(\tau)
    v^k(0)\right>=\frac {1}{3} \left< {\vec v}^2\right> \delta^{ik}
    e^{-\Gamma_x |\tau|},
\end{equation}
where $\Gamma_x$ is the relaxation rate which is supposed to be
approximately constant on the relaxation time scale $1/\Gamma_x$.  It
relates to the spatial diffusion coefficient $D$ via Einstein's
relation

\begin{equation}\label{Einstein}
    D=\frac{1}{3}\int^\infty_0 \di \tau \left< {\vec v}(\tau)
    {\vec v}(0)\right>=\frac{1}{3\Gamma_x} \left< {\vec v}^2\right>\,.
\end{equation}
Compared to the infra--red divergent quasi--free result $\propto
1/\omega^2$ (c.f. eq. (\ref{MIQF}) below) this form of the correlation
renders the photon self energy
\begin{equation}\label{Mrelax} -\ii\Pi^{-+}_{\mathrm{cl}}(\omega,{\vec
    q}=0)= 4\pi e^2\rho_0\left< v^iv^k\right>
    \frac{2\Gamma_x}{\omega^2+\Gamma_x^2}= 4\pi e^2\rho_0\frac{
     2 D\Gamma_x^2}{\omega^2 +\Gamma_x^2}\delta^{ik}
\end{equation}
regular at four momentum $q=0$. It is determined by mesoscopic
transport properties, namely by the diffusion coefficient $D$ and
relaxation rate $\Gamma_x$; $\rho_0$ is the spatial density of the
charged particles.

Both $f({\vec x},{\vec v},t)$ and the autocorrelation function can be
obtained in closed form, if the time evolution of $f$, and the
propagation of fluctuations $\delta f$ are governed by a standard
(non-relativistic) diffusion process (Fokker--Planck equation)

\begin{equation}\label{FP}
    \frac{\partial}{\partial t} f({\vec x},{\vec v},t)=
    \left({D\Gamma_x^2}\frac{\partial^2}{\partial {\vec v}^2}
    +\Gamma_x \frac{\partial}{\partial {\vec v}}{\vec v}-{\vec
    v}\frac{\partial}{\partial {\vec x}}\right) f({\vec x},{\vec v},t).
\end{equation}
In the equilibrium limit ($t\rightarrow\infty$) the distribution
attains a Maxwell-Boltzmann form
\begin{equation}\label{feq} f_{eq}({\vec x},{\vec
    v})=\rho_0 f_{eq}({\vec v})=\rho_0 \left(2\pi
    D\Gamma_x\right)^{-3/2} \exp\left[-\frac{v^2}{2D\Gamma_x}\right]
    =\frac {m^3}{(2\pi)^3}e^{-(\epsilon(v)-\mu)/T}\; ,
\end{equation}
where $T=m\left< {\vec v}^2\right>/3=mD\Gamma_x$  and $\mu$ are the
equilibrium temperature and chemical potential and $\epsilon(v)$ is
the energy of the particle.

At $\tau=0$ we consider an initial fluctuation $\delta f({\vec
x},{\vec v},\tau=0)=\delta^3({\vec x})\delta^3({\vec v}-{\vec
v}_0)$. Its propagation in the equilibrated matter is also governed by
the Fokker Planck equation (\ref{FP}). By a Gaussian ansatz for the
Fourier transform of this fluctuation $\delta{\tilde f}({\vec q},{\vec
y},\tau)=\int \di^3x\di^3v f({\vec x},{\vec v},\tau) \exp[-\ii{\vec
q}{\vec x}+\ii{\vec y}{\vec v}]$ the time-dependence can be obtained
in closed form as

\begin{eqnarray}\label{ftilde} &&\delta {\tilde f}({\vec q},{\vec
    y},\tau)=\exp\left[ -A+\ii{\vec B}{\vec y}-Cy^2\right],
    \quad\mathrm{where}\cr
    &&C=\frac{D\Gamma_x}{2}\left(1-e^{-2\Gamma_x\tau}\right),\quad {\vec
    B}= {\vec v}_0e^{-\Gamma_x\tau}-\ii{\vec
    q}D\left(1-e^{-\Gamma_x\tau}\right)^2,\\ &&A=\ii{\vec
    q}\int_0^\tau\di \tau'{\vec B(\tau')}
    =\frac{Dq^2}{2\Gamma_x}\left[2\Gamma_x \tau-e^{-2\Gamma_x
    \tau}+4e^{-\Gamma_x \tau} -3\right]-\frac{\ii{\vec
    q}\;{\vec v}_0}{\Gamma_x}\left( e^{-\Gamma_x \tau}-1\right).\nonumber
\end{eqnarray}
This fluctuation $\delta f$ is the conditional probability which
determines the time-depen\-dence of the current autocorrelation
function. With four vectors $\{v^\mu\}=\{1,{\vec v}\}$ and
$\{B^\mu\}=\{1,{\vec B}\}$ one can express the full correlation tensor
in the mixed $\tau,{\vec q}$ representation as
\begin{eqnarray}\label{Pidiff}
    \hspace*{-1cm}&-\ii&\Pi_{\mathrm{cl}}^{\mu\nu -+}(\tau, {\vec
    q})=4\pi\int \di^3  x e^{-\ii{\vec q}{\vec x}}
    \left<j^\nu({\vec
    x},\tau)j^\mu({\vec 0},0)\right>\cr &=&4\pi e^2\rho_0 \int\di^3v
    f_{eq}({\vec v})
    v^\mu B^\nu e^{-A}
    =4\pi e^2\rho_0\left< v^\mu B^\nu e^{-A}\right>_{eq}\cr
    &=&4\pi e^2\rho_0\exp\left\{-\frac{D\,{\vec q}^2}{\Gamma_x}
    \left(\Gamma_x|\tau|+e^{-\Gamma_x|\tau|}-1\right)\right\}\\
    &&\times\left\{\begin{array}{ll}
    \left\{\left<  v^\mu v^\nu \right>_{eq}
    e^{-\Gamma_x|\tau|} -D^2 q^\mu q^\nu\left(e^{-\Gamma_x|\tau|}-1\right)^2
    \right\}&\for \mu,\nu\in\{ 1,2,3\}\cr
    \ii q^\nu D\left(e^{-\Gamma_x|\tau|}-1\right){\mathrm sign}(\tau)
   &\for \mu=0,\;\nu\in\{1,2,3\}\cr
    1&\for \mu=\nu=0 \end{array}\right.\nonumber
\end{eqnarray}
with $A$ and ${\vec B}$ as a function of $\tau$, ${\vec q}$ and ${\vec
v}_0={\vec v}$ from (\ref{ftilde}). Here the ensemble average
\mbox{$\left<\dots\right>_{eq}$} over the equilibrium distribution
$f_{eq}$ keeps only even moments of ${\vec v}$ with \mbox{$\left<{\vec
v}^2\right>= 3D\Gamma_x$}. The result complies exactly with current
conservation, i.e. one verifies $\partial_0\Pi^{0\nu}+\ii
q_k\Pi^{k\nu}\equiv 0$.

For transverse photons terms proportional to $q^\mu q^\nu$ drop.  The
corresponding spatial part of the tensor is shown in fig. 1, right
part. This correlation function decays exponentially as $\sim
e^{-\Gamma_x\tau}$ at ${\vec q}=0$, and its width further decreases
with increasing momentum $q=|{\vec q}|$ due to the increase in spatial
resolution. The left part shows the corresponding density-density
correlation ($\mu=\nu=0$), which decays only for non-zero momentum, due
to charge conservation.

\noindent
\epsfig{file=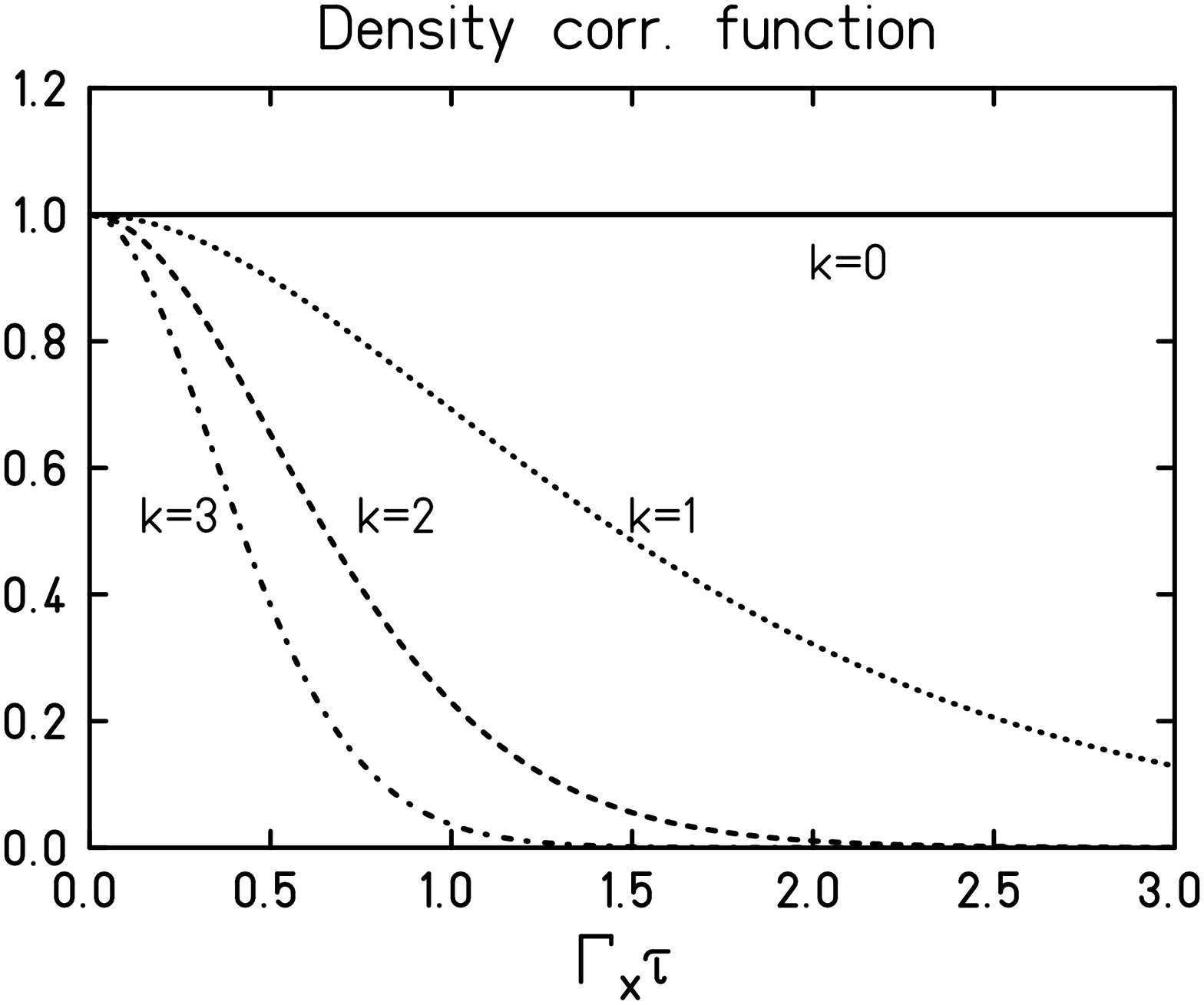,width=7.2cm,height=6cm} \hspace*{4mm}
\epsfig{file=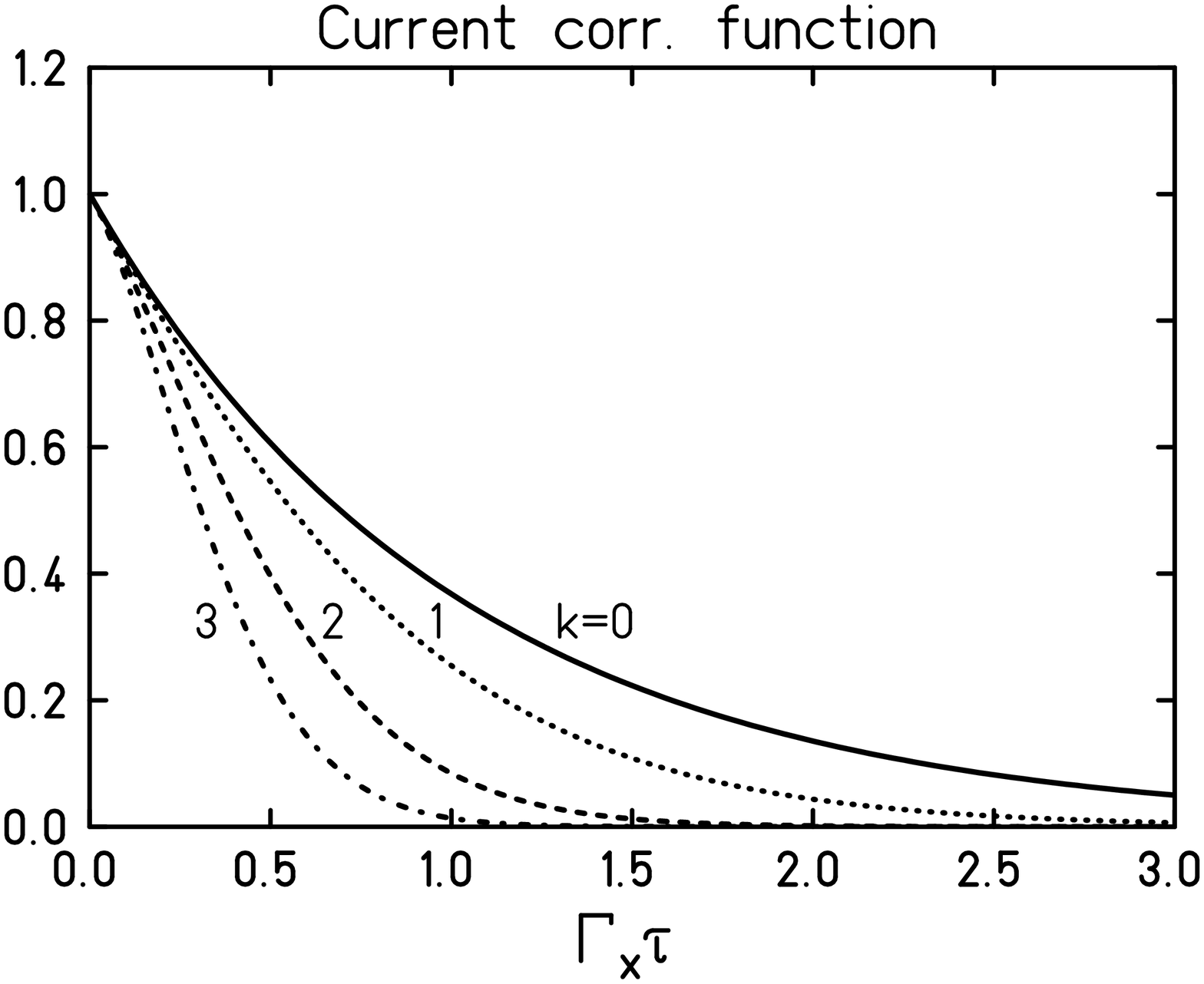,width=7.2cm,height=6cm} \\
\parbox[t]{15cm}{\small Fig. 1: Density-density and current-current
correlation functions, $-\ii\Pi_{\mathrm{cl}}^{00\;-+}(\tau,{\vec q}\,)$
and $-\ii\Pi_{\mathrm{cl}}^{11\;-+}(\tau,{\vec q}\,)$, normalized to the
values at $\tau=0$ as a function of time $\tau$ (in units of
$1/\Gamma_x$) for different values of the photon momentum
$q^2=3k^2\Gamma_x^2/{<v^2>}$ with $k=0,1,2,3$.}

The remaining time Fourier transformation gives the $\omega,{\vec
q}$-dependence of the photon self-energy. It can be expressed in terms
of the incomplete gamma function.  Straightforward expansion in powers
$\left\{(D{\vec q}^2/\Gamma_x)e^{-\Gamma_x|\tau|}\right\}^n$ leads to

\begin{eqnarray}\label{Mclres}
    -\ii\Pi_{\mathrm{cl}}^{-+}(\omega,{\vec q}\,)
    &=&4\pi e^2 \rho_0\left< v^iv^k\right>_{eq}
     \exp\left[{D\,{\vec q}^2}/{\Gamma_x}\right]\cr &&\times
     \sum_{n=0}^\infty \frac{1}{n!}
     \left(\frac{-D\,{\vec q}^2}{\Gamma_x}\right)^n
     \frac{2(n+1)\Gamma_x+2D\,{\vec q}^2}
     { \left((n+1)\Gamma_x+D\,{\vec q}^2\right)^2+\omega^2}
\end{eqnarray}
for transverse photons. Since the correlation functions are properly
determined from the time structure of the source, they comply with the
$n=2$ (and $n=1$) sum-rule constraints.

There are two limiting cases where simpler analytical forms can be
obtained: i) at small momentum transfers where eq. (\ref{Mclres}) can
be expanded in powers of ${\vec q}$ and rewritten as to provide a
propagator type form and ii) for large momentum transfers where from the
$\exp\{\dots\}$ part in (\ref{Pidiff}) a short time Gaussian
behavior emerges. Thus, for  small momentum transfers one finds
\begin{eqnarray}\label{Mclres0}
   \lim_{\left< v^2\right>{\vec q}^2\ll \Gamma_x^2}
    &&\left\{-\ii\Pi_{\mathrm{cl}}^{-+}(\omega,{\vec q}\,)\right\}
    =4\pi e^2 \rho_0\left< v^iv^k\right>_{eq}
     \frac{2\Gamma_x}{\Gamma_x^2+\omega^2+
     \frac{2\Gamma_x^2-4\omega^2}{3(4\Gamma_x^2+\omega^2)}
     \left<{\vec v}^2\right>{\vec q}^2  }
          \nonumber\\
    &&\approx4\pi e^2 \rho_0\left< v^iv^k\right>_{eq}
     \frac{2\Gamma_x}{\Gamma_x^2+\omega^2+\left<{\vec v}^2\right>
    {\vec q}^2/6} \for \omega\ll\Gamma_x
\end{eqnarray}
which generalizes the relaxation result (\ref{vrelax},\ref{Mrelax}) to
finite ${\vec q}$.
On the other hand for large  momenta  one realizes that
    \begin{eqnarray} \label{Pidifflq}
    \lim_{D\,{\vec q}^2\gg\Gamma_x}
    \left[-\ii\Pi_{\mathrm cl}^{-+}(\tau,{\vec q}\,)\right] &=&4\pi
    e^2\rho_0\left< v^iv^k\right>_{eq}
    \exp\left[-D\,{\vec q}^2\Gamma_x\tau^2/2\right]\;
    ,\quad\mbox{and therefore}\cr \lim_{D\,{\vec q}^2\gg\Gamma_x}
    \left[-\ii\Pi_{\mathrm cl}^{-+}(\omega,{\vec q}\,)\right] &=&4\pi
    e^2\rho_0\left< v^iv^k\right>_{eq} \sqrt{\frac{2\pi}{D\,{\vec
    q}^2\Gamma_x}} \exp\left\{-\frac{\omega^2}{2D\,{\vec
    q}^2\Gamma_x}\right\}\\
    &=&4\pi
    e^2\left< v^iv^k\right>_{eq}
    \frac{m^2T}{2\pi|{\vec q}|}
    \exp\left\{-\left(\mbox{$\frac{m}{2}$}\omega^2/|{\vec
    q}|^2-\mu\right)/T \right\}\; ,\nonumber
\end{eqnarray}
where obviously the essential contributions come from velocities which
satisfy the Cherenkov condition $|{\vec v}|\approx \omega/|{\vec q}|$.
This limit is independent of the relaxation rate $\Gamma_x$ and
coincides with the quantum one-loop diagram result in the
corresponding large $|{\vec q}|$ limit, as we shall see in sect. 6.

Although the above expressions give the exact solution of the
mathematical problem posed in this section, its physical
interpretation has to be done with some care for the following
reason. The equilibrium source distribution contains velocity
components that exceed the speed of light. Therefore for the physical
result mistakes of the order of $\exp[-3/(2\left<
v^2\right>)]=\exp[-m/(2T)]$ are expected. This restricts the
application to non-relativistic sources and for large ${\vec q}$ to
space-like photons, where $|{\vec q}|\gg\omega $.

For systems with given fixed mean-square velocity $\left< v^2\right>
=const.$ the exact classical on-shell rate (\ref{dN}) at $|{\vec
q}|=\omega$ evidently scales as a function of $\omega/\Gamma_x$. It
properly vanishes at $\omega=0$ and at infinity. It is important to
note that the rate has an upper bound of $\approx
\frac{2}{3\pi}e^2\rho_0\left< v^2\right>$, and indeed attains its
maximum value around $\omega\approx\Gamma_x$, which is collision-rate
independent. For simplicity we quote the closed form obtained in the
non-relativistic limit (\ref{Mclres0}), which coincides with the
dipole limit. There
\begin{equation}\label{dNmax}
    \frac{\di^5 {n_{\gamma}}}{\di^3 x\di\omega\di t}\approx
    \mbox{$\frac{4}{3\pi}$}e^2\rho_0\left< v^2\right>
    \frac{\omega/\Gamma_x}{1+(\omega/\Gamma_x)^2}
    \rightarrow\mbox{$\frac{4}{3\pi}$}e^2\rho_0\left< v^2\right>
    \left\{\begin{array}{ll}
    \Gamma_x/\omega&\quad\mbox{ for }\omega\gg\Gamma_x\\
    \mbox{$\frac{1}{2}$}&\quad\mbox{ for }\omega=\Gamma_x\\
    \omega/\Gamma_x&\quad\mbox{ for }\omega\ll\Gamma_x .
    \end{array}\right.
\end{equation}
One realizes that the ultra-violet part of the spectrum
$\omega\gg\Gamma_x$ behaves as intuitively expected, fig. 2: the rate
grows proportional to the relaxation rate, until it saturates around
$\omega\approx\Gamma_x$. For the soft part $\omega\ll\Gamma_x$,
however, the rate becomes inversely proportional to the collision
rate! The higher the collision rate the more suppressed the
spectrum. In order to illustrate the non-perturbative character of
this soft behavior supposes $\Gamma_x\propto\ g^2$, where $g$ is the
strong coupling constant of the source system. One sees that indeed
the low-$\omega$ part with $\Pi\propto e^2\omega/\Gamma_x\propto
e^2\omega/g^2$ represents a genuine non-perturbative result in $g$,
while the large $\omega $-part, where $\Pi\propto e^2g^2/\omega$, is
well described perturbatively.

\parbox[t]{7.2cm}{\epsfig{file=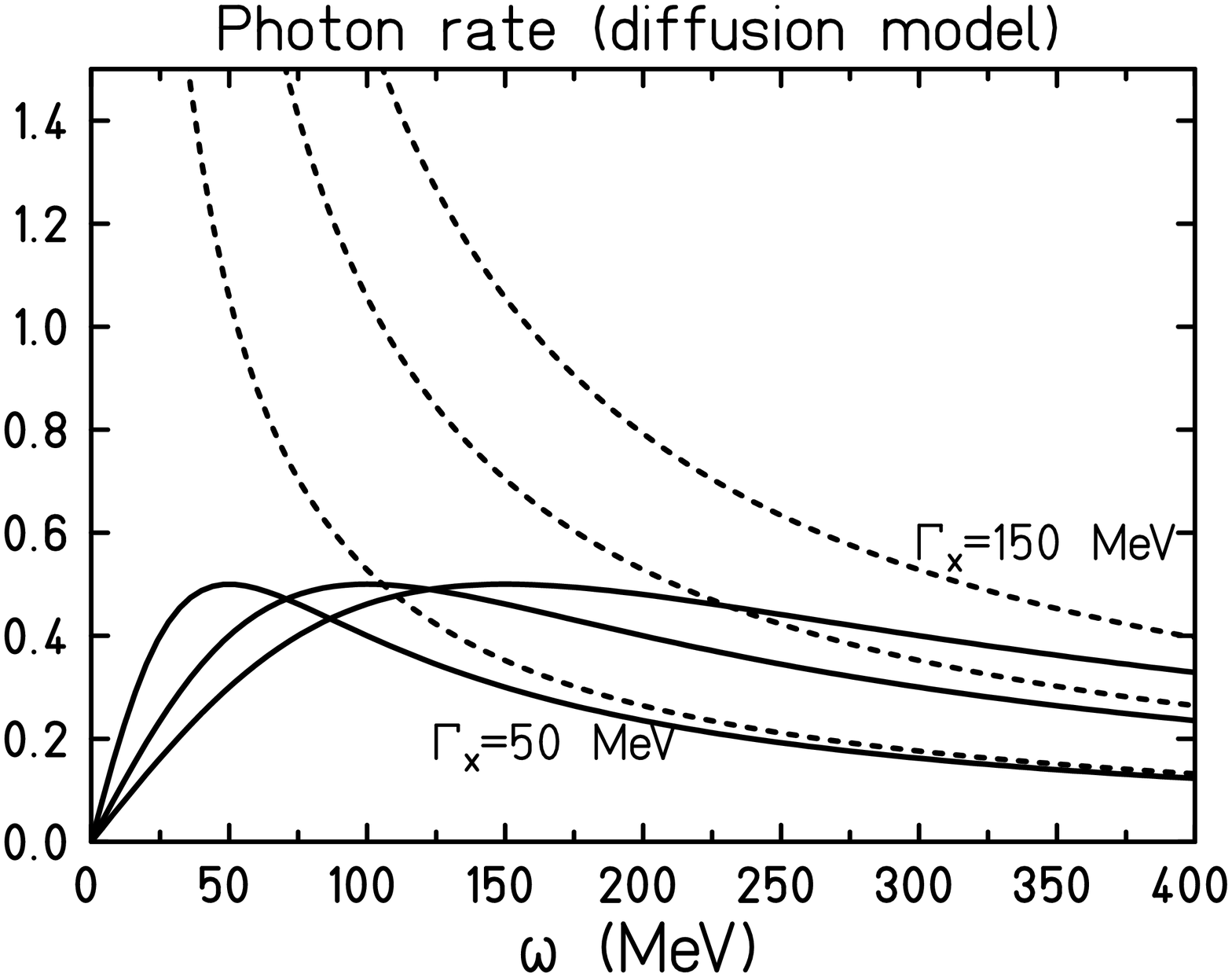,width=7.2cm,height=6cm} \\
{\small Fig.~2: Rate of real photons $\di
N/(\di\omega\di t)$ in units of $4\pi e^2\left<{\vec v}^2\right>/3$
for a non-relativistic source for $\Gamma_x=$50,100,150 MeV; for
comparison the IQP results (dashed
lines) are also shown.}}
\hspace*{4mm}\parbox[t]{7.2cm}
{\epsfig{file=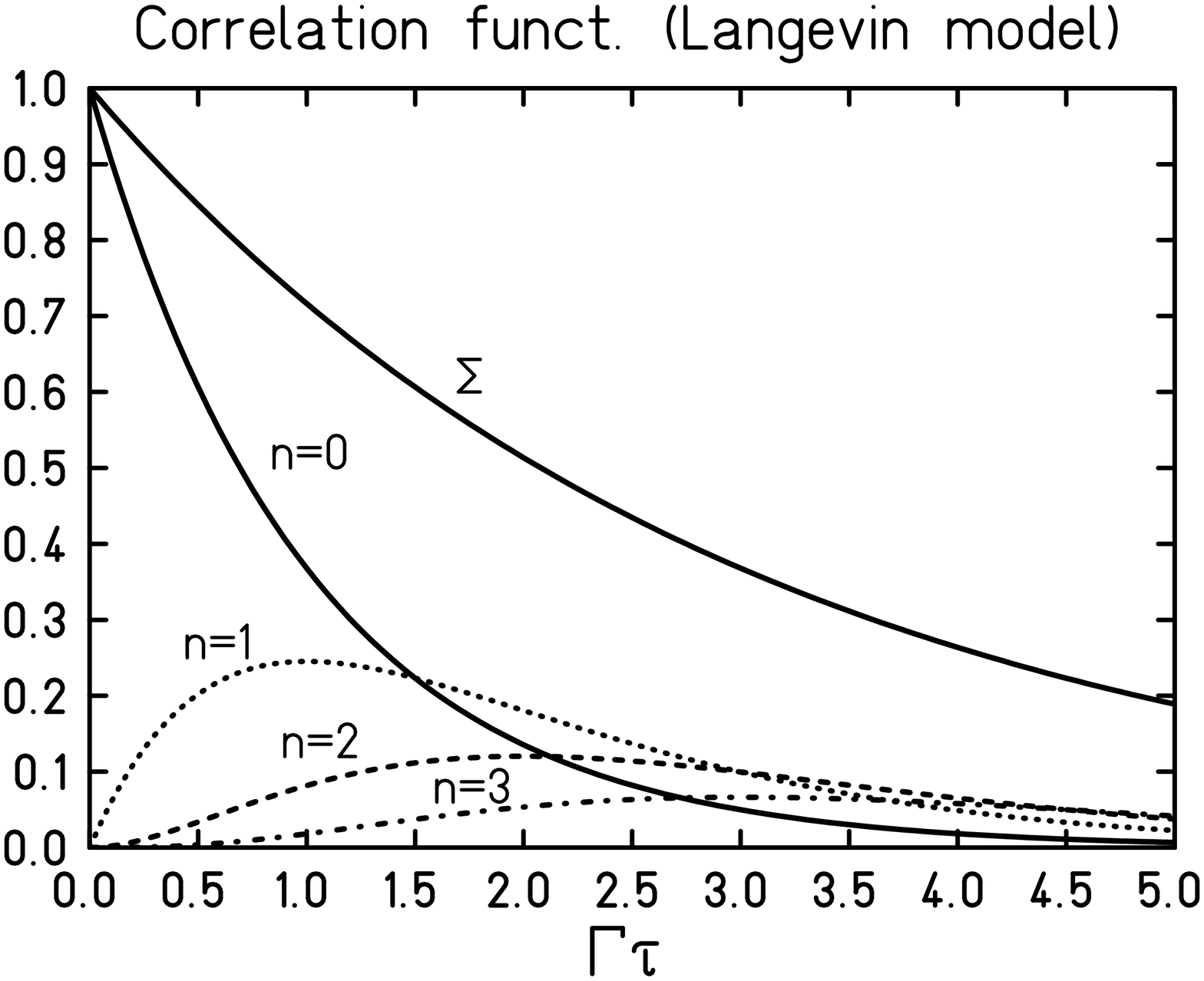,width=7.2cm,height=6cm}
\\ {\small Fig.~3: Current correlation function for the first terms
$n=0,1,2,3$ of the Langevin result, eq. (\ref{Apoisson}) and the total
sum ($\Sigma$) for the case that\\ $\Gamma_x/\Gamma=\left< ({\vec
v}_m-{\vec v}_{m+1})^2 \right>/(2\left< {\vec v}^2 \right>)=1/3$.}}
\\[-4mm]

\subsection{Microscopic Langevin Process}

For a later comparison with quantum diagrams in sect. 4 we should look
into the corresponding microscopic picture of classical
propagation. There one can consider a classical random process
(Langevin process), where hard scatterings occur at random with a
constant mean collision rate $\Gamma$. These scatterings consecutively
change the velocity of a point charge from ${\vec v}_m$ to ${\vec
v}_{m+1}$ to ${\vec v}_{m+2}$, $\dots$ (in the following subscripts
$m$, $n$, and $l$ refer to the collision sequence, while superscripts
$i,k\in \{1,2,3\}$ specify the spatial components of vectors and the
self-energy tensor). In between scatterings the point charge moves
freely. For such a multiple collision process some explicit results
can be given. They all refer to the case of vanishing photon momentum
${\vec q}=0$ (dipole approximation) and therefore apply to
non-relativistic sources where $\left< v^2\right>{\vec
q}^2\ll\Gamma_x^2$ or to dilepton production, for example, since only
the time structure is well known in this case, while the space
structure would require an integration of the random classical paths.

For such a Langevin process the modulus of the autocorrelation function
takes a simple Poissonian form (fig. 3)
\begin{eqnarray}\label{Apoisson}
   -\ii\Pi_{\mathrm{cl}}^{-+}(\tau,{\vec q}=0)&=&4\pi e^2\rho_0
   \left<v^i(\tau)v^k(0)\right>\cr &=&4\pi
   e^2\rho_0e^{-|\Gamma\tau|}\sum_{n=0}^\infty
   \frac{|\Gamma\tau|^n}{n!} \left< v^i_m v^k_{m+n}\right>_m\, .
\end{eqnarray}
Here $\left<\dots\right>_m$ denotes the average over the discrete
collision sequence $\{m\}$. This form, which one writes down
intuitively, directly includes what one calls {\em damping} and
therefore corresponds to a resummation description in the quantum
case. The corresponding perturbation theory result is obtained through
an expansion in powers of $\Gamma$
\begin{equation} \label{Apower}
    -\ii\Pi_{\mathrm{cl}}^{-+}(\tau,{\vec q}=0)=4\pi e^2\rho_0
    \sum_{n=0}^\infty\frac{|\Gamma\tau|^n}{n!}\sum_{l=0}^n (-1)^k
    {n\choose l} \left<v_m^i v_{m+l}^k\right>_m,
\end{equation}
which for dimensional reasons is also a power series in $|\tau |$
in this case.  If the $\left<v^i_m v^k_{m+n}\right>$ expectation
values are replaced by unity in (\ref{Apoisson}) or (\ref{Apower}),
one obtains the $00$-component of $\Pi$ which becomes constant in time
in line with (\ref{timewardid}).

The time Wigner transform of (\ref{Apoisson}) determines
the $\omega$-spectrum at vanishing $\vec q$

\begin{equation} \label{Mtau}
    -\ii\Pi_{\mathrm{cl}}^{-+}(\omega,{\vec q}=0)
    =4\pi e^2\rho_0 \sum_{n=0}^\infty
     \left< v_m^i v_{m+n}^k\right>_m\Re\left\{
    \frac{2\Gamma^n \left\{ (\Gamma+\ii\omega )^{n+1}\right\} }
    {(\omega^2+\Gamma^2)^{n+1}}\right\} .
\end{equation}
This is a genuine {\em multiple collision} description for the photon
production rate in completely regular terms due to the
\mbox{$(\omega^2+\Gamma^2)^n$} form of all denominators. Each term is
regular, since right from the beginning one accounts for the damping
of the source particle because of its finite mean time $1/\Gamma$
between collisions. The result (\ref{Mtau}) still accounts for the
{\em coherence} of the photon field, now expressed through the
correlations $\left<{\vec v}_m{\vec v}_{m+n}\right>_m$ arising from
the sequence of collisions. Note in particular, that, although the
total expression is positive, the $n>0$ terms can be negative since
they describe the interference of the radiation arising from different
propagation segments of the source particle.  Thus, the terms in
(\ref{Mtau}) define partial rates, which later (sect. 5.2) will be
associated with specific self energy diagrams.

As already mentioned, the ${\vec q}$-dependence of the self energy
cannot be given in closed form in general apart from the $n=0$ term
(c.f. with $n=0$ term from eq. (\ref{Mtau}))
\begin{equation}\label{Mtaun0}
-\ii\Pi_{\mathrm{cl}}^{-+}(\omega,{\vec q})
    \approx 4\pi e^2\rho_0
     \left< \frac{2\Gamma \,v_m^i v_m^k}{(\omega -{\vec q}{\vec v})^2
    +\Gamma^2}\right>_m\,.
\end{equation}
It shows the typical Cherenkov enhancement at $\omega={\vec q}{\vec
v}$.  At this level one may be tempted to associate this ($n=0$) term
with the relaxation ansatz result (\ref{Mrelax}). This however is only
true if $<{\vec v}_m\;{\vec v}_{m+n}>_m=0$ for $n\ne 0$, an
approximation recently used in refs. \cite{Cleymans}. In the general
case velocity correlations between successive scatterings exist, and
there will be a difference between the microscopic mean collision rate
$\Gamma$ and the mesoscopic relaxation rate $\Gamma_x$. Still, for
systems, where the velocity is degraded by a constant fraction
$\alpha$ per collision, such that $\left< {\vec v}_m\cdot{\vec
v}_{m+n} \right>_m= \alpha^n\left< {\vec v}_m\cdot{\vec v}_{m}
\right>_m$, one can resum the whole series in (\ref{Mtau}) and thus
recover the relaxation result (\ref{Mrelax}) at ${\vec q}=0$. The
macroscopic rate $\Gamma_x$ is then determined by the microscopic
scattering properties through $\Gamma_x=(1-\alpha)\Gamma$, or $2\left<
({\vec v}_m)^2 \right>_m\Gamma_x=\left< ({\vec v}_m-{\vec v}_{m+1})^2
\right>_m\Gamma$.  This clarifies that the diffusion result
(\ref{Mclres}) represents a resummation of the random multiple
collision result.

The following relations show different reformulations and limits of
the Langevin result (\ref{Mtau}). For instance the invariance of
(\ref{Mtau}) is not directly visible, since absolute velocities
enter. Still the perturbation expression (\ref{Apower}) can be
rewritten, such that except for the zero order term, which drops out
in the Fourier transform, only velocity differences appear

\begin{eqnarray} \label{Ainv}
   &-\ii\Pi_{\mathrm{cl}}^{-+}(\tau,{\vec q}=0)
    =4\pi e^2\rho_0 \left\{ \left< v_m^i v_m^k\right>_m
    -\frac{|\Gamma\tau|}{2}\left< (v_m^i- v_{m+1}^i)(v_m^k-
    v_{m+1}^k)\right>_m \right. \cr
    &\left. -\sum_{n=2}^\infty\frac{|\Gamma\tau|^n}{n!}\sum_{l=0}^{n-2}
    (-1)^k {n-2\choose l} \left<(v_m^i-v_{m+1}^i)(v_{m+l+1}^k-v_{m+l+2}^k)
\right>_m\right\} .
\end{eqnarray}
Terms of lowest odd order in $|\tau|$ determine the asymptotic large
$\omega$ (ultra violet) behavior of the spectrum

\begin{eqnarray} \label{Muv}
    &\lim_{\omega\to\infty}
    \left[-\ii\Pi_{\mathrm{cl}}^{-+}(\omega,{\vec q}=0) \right]
     =\frac{4\pi e^2\rho_0}{\Gamma}\left\{
    \frac{\Gamma^2}{\omega^2}
   \left<(v_m^i-v_{m+1}^i)(v_m^k-v_{m+1}^k)\right>_m  \right. \cr
    &\left. +
    \sum_{n=2}^\infty\left(\frac{\Gamma}{\omega}\right)^{2n}\;
    \sum_{l=0}^{2n-3}
    (-1)^l {2n-3\choose l} \left<(v^i_m-v^i_{m+1})(v^k_{m+l+1}-v^k_{m+l+2})
    \right>_m\right\} .
\end{eqnarray}
Apart from the mean collision time $\approx 1/\Gamma$ this is an expansion
in powers of $(\Gamma/\omega)^2$ and therefore represents the
perturbation expansion result for the classical source ($\Gamma$
representing the interaction, while $1/\omega$ relates to the
intermediate propagator). This perturbation expansion (\ref{Muv}) is
interesting since it already displays the main problem: While for
$\omega\gg \Gamma$ the series converges, if higher order correlations
cease sufficiently fast, there is no hope to ever recover the correct
result (\ref{Mtau}) for $\omega\ll\Gamma$. This is so, since i) this
series is necessarily divergent (it has to recover
$1/(\omega^2+\Gamma^2)$ by a power series in $\Gamma$), but also it
misses the knowledge on the even powers in $\tau$ in eq. (\ref{Ainv})
which essentially determine the soft behavior.

The first term in (\ref{Muv}) represents the {\em incoherent
quasi--free} production rate which for finite ${\vec q}$ at given
polarization $\eta$ reads

\begin{equation}\label{MIQF}
   -\ii\eta_\mu\eta_\nu\Pi^{\mu\nu-+}_{\mathrm{IQF}}
   (\omega,{\vec q})
   =4\pi{e^2}\rho_0\Gamma\left<
   \left|\frac{\eta k_m}{qk_m}-\frac{\eta k_{m+1}}{qk_{m+1}}
   \right|^2\right>_m\,.
\end{equation}
It carries the known divergence at the soft point $q=0$, c.f. dashed
lines in fig. 2 and 4. In conclusion: {\em the commonly used IQF
prescription fails for soft particle production. }

\subsection{Finite Size Corrections}
For systems of finite spatial extension and conserved currents one can
consider the $n=0$ sum rule. It demands that $\Pi^{\mu\nu-+}(q;x)$ has
to vanish at least quadratically with $q\rightarrow 0$.  This property
survives in the classical limit, where formally $\hbar\rightarrow 0$
and the spectrum becomes continuous.  Thus, the term of zero order in
$\omega$ given by $\int_{-\infty}^\infty \left<({\vec v}(\tau))({\vec
v}(0))\right> \di \tau$ has to drop and the ensuing low energy part of
the spectrum (\ref{Mtau}) starts quadratically in $\omega$

\begin{eqnarray} \label{Mir} \lim_{\omega\to 0} &\left[-\ii\Pi_{\mathrm
    cl}^{-+}(\omega,{\vec q}=0)\right]=\cr
    &-\frac{4\pi e^2\rho_0}{\Gamma}\left\{\frac{\omega^2}{\Gamma^2}
    \sum_{n=0}^\infty (n+1)(n+2)\left< v^i_mv^k_{m+n} \right>_m
    + {\cal O}\left(
    \frac{\omega^4}{\Gamma^4}\right)\right\} .
\end{eqnarray}
The simple relaxation ansatz (\ref{vrelax}) does not fulfill this
finite size condition, since it ignores long term anti-correlations on
the scale of some recurrence time $1/\Gamma_{rec}$. That is the time,
where on the mean the center of charges returns to the same
point\footnote{not to be confused with the Poincare recurrence time,
which is of no relevance here.}. This defect of the relaxation ansatz
may be cured by a more general form which includes such an
anti-correlation, e.g.
\begin{eqnarray}\label{vrec}
    \left< {\vec v}(\tau) {\vec v}(0)\right>&=&
    \left< {\vec v}^2\right> \left\{e^{-\Gamma_x
    |\tau|}-\frac{\Gamma_{rec}^2}
    {\Gamma_x}|\tau|e^{-\Gamma_{rec}|\tau|}\right\}\; ;\cr
    -\ii\Pi_{\mathrm{cl}}^{-+}(\omega,{\vec q}=0) &=&4\pi e^2\rho_0
    \left< v^iv^k\right>\frac{2}{\Gamma_x}
    \left\{\frac{\Gamma_{x}^2}{\omega^2+\Gamma_x^2}+
    \frac{\Gamma_{rec}^2 (\omega^2-\Gamma_{rec}^2)}
    {(\omega^2+\Gamma_{rec}^2)^2}
    \right\} .
\end{eqnarray}
The extra parameter $\Gamma_{rec}$ can be determined such that the
spectrum fulfills the $n=0$ dipole sum-rule (\ref{sumrule}). For
larger systems one infers that $\Gamma_{rec}\approx
2\left<v^2\right>/(\left<x^2\right>\Gamma_x)$, where $\left<x^2\right>$
is the mean-square extension.

For small systems both time scales become comparable and the self
energy tensor attains the form

\begin{equation}
    -\ii\Pi_{\mathrm{cl}}^{-+}(\omega,{\vec q}=0) =4\pi e^2
    \rho_0\left< v^iv^k\right>
    \frac{4 \Gamma_x\omega^2}{(\omega^2+\Gamma_x^2)^2} \qquad{\mathrm
    for}\qquad\Gamma_x=\Gamma_{rec}.
\end{equation}
This form has been found in the one-dimensional model of
ref. \cite{KnollLenk}, where due to the dominance of back scattering
both time scales merge.

\noindent
\parbox{7.cm}{ \small Fig.~4: Current correlation function
$-\ii\Pi^{11-+}(\omega,{\vec q}=0)$ for a random collision sequence
limited to a finite size in space such that $\left<x^2\right>\simeq
10\left<v^2\right>/\Gamma^2$ and $\Gamma_x=0.8\Gamma$: full line from
Monte Carlo calculation of the phase integral (\ref{M2cl});
dashed-dotted line from the analytic Langevin result (\ref{Mtau});
dotted line includes finite size corrections (\ref{vrec}); dashed line
from quasi-free scattering prescription.}\hspace*{4mm}
\parbox{7.2cm}{\epsfig{file=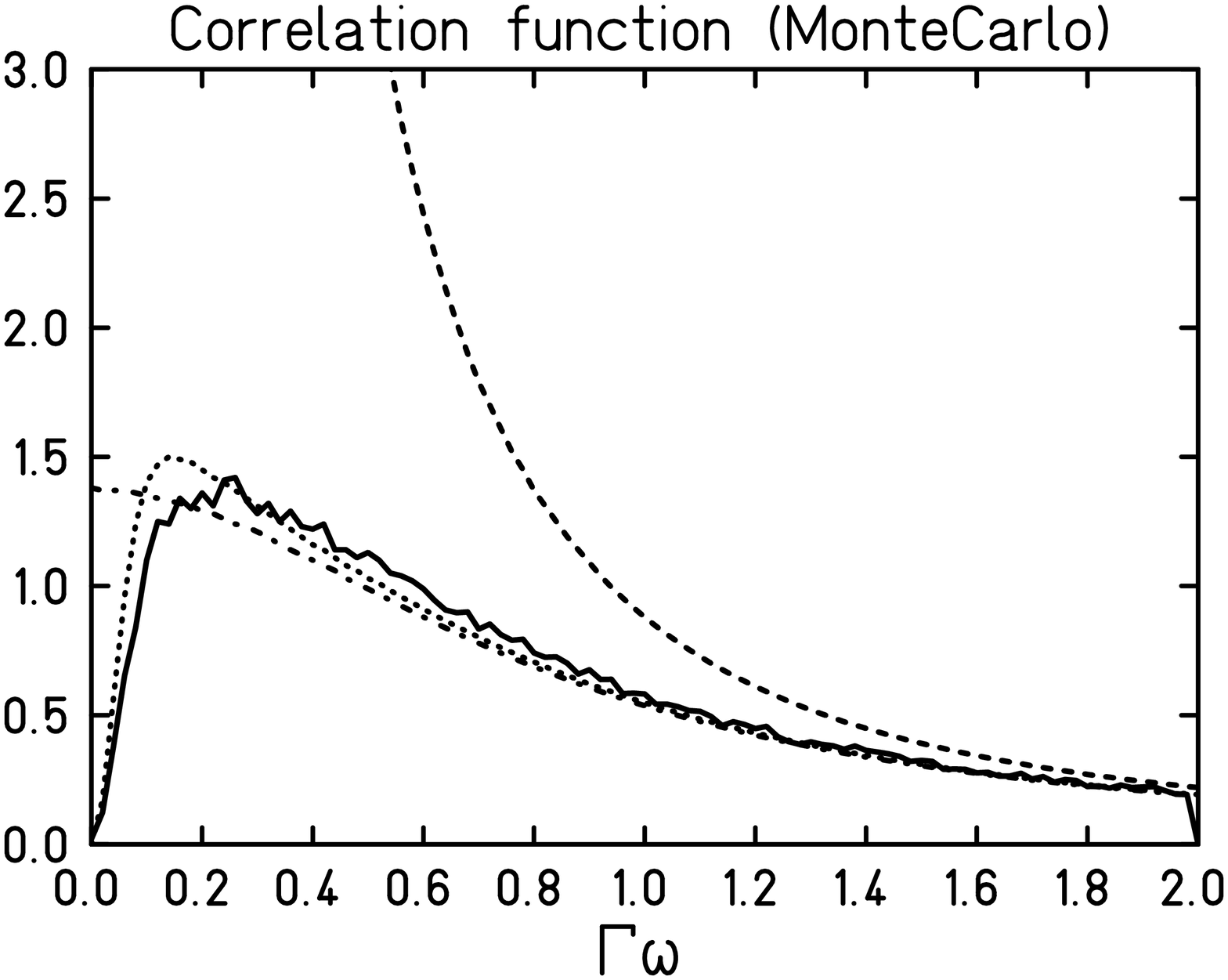,width=7.2cm,height=6cm}\\}
\\[-4mm]

\subsection{Comparison with Monte Carlo Evaluation of Amplitudes}
For illustration we like to present a simple model result and compare
it to a Monte Carlo method, where amplitudes are calculated by
considering the phase of the photon field along the classical orbits.
Thus, one evaluates
\begin{equation}\label{M2cl}
\int \di t {\vec v}(t)\exp[\ii\omega t-\ii{\vec q}{\vec x}(t)]
\end{equation}
along the random straight sections of the classical paths in the
cascade model. This method also permits to illustrate the finite size
corrections discussed above. The dotted line in fig. 4. shows the
Monte Carlo phase integral result (\ref{M2cl}) from an oriented random
walk, compared to the simple relaxation result (\ref{Mrelax}), dashed
line. The example has the property that the motion is limited to a
finite space with $\left<x^2\right>\simeq 10\left<v^2\right>/\Gamma^2$
and $\Gamma_x=0.8\Gamma$. The sharp dip at $\omega=0$ and the little
over shoot around $\omega\sim0.3\Gamma$ are due to the finite size of
the system. The full line gives the relaxation result including the
finite size corrections (\ref{vrec}).

Evidently the Monte Carlo method is a highly unreliable due to the
strong cancelations of terms that are randomly generated. The
precision in fig. 4 is obtained with $200$ cascade runs where each
path has about $10^3$ collisions (simulation codes have by far less
statistics!).  The analytical result (\ref{Mtau}) has significant
computational advantages, provided the random process is of this
form. For the same precision already a single representative path with
about $10^3$ collisions is sufficient, while for the relaxation ansatz
one only has to determine the relaxation rate, i.e. a simple moment in
time.

\subsection{Infra-red Divergences, Current Conservation, Gauge
Invariance and Identities}

In the above multiple collision description {\em with damping}
(\ref{Apoisson}) all terms have a finite range in time. Thus, they all
are void of infra-red divergences and so is any limited sum of terms
of them.  This is the desirable feature that we are aiming at. In
addition the example above illustrates that in many cases only a few
rescattering terms are necessary in order to properly recover the
correct result for the transverse part of the correlation tensor both
at small and large $\omega$. In fact the number $n_x$ of required
rescattering terms is given by $n_x\sim \Gamma/\Gamma_x$.

The picture is more subtile for the longitudinal components of the
tensor, since some particular integrals are conserved (time
independent) as they related to the total charge,
c.f. eq. (\ref{timewardid}), sect. 2.1. Thus $-\ii\Pi^{00-+}({\vec
q}=0,\tau)$, c.f. fig. 1, has to be constant. For the Langevin case
this identity holds, since
\begin{equation}
\sum_{n=0}^{\infty} \frac{(\Gamma\tau )^n}{n!}e^{-\Gamma\tau}\equiv 1.
\end{equation}
However, for any finite number of such terms this identity can never be
recovered exactly for all times, since the relaxation
time of total charge is infinite!

Quite often one is interested in a solution of the problem up to a
maximal time $t_{max}$, and one likes to request charge conservation
on the correlator level to be maintained only within this time span or
within the corresponding $\omega$-range limited to $\omega
>\omega_{min}=1/t_{max}$. With this limitation a finite number of
terms $n_x\sim\Gamma t_{max}\sim\Gamma/\omega_{min}$ is required for a
proper description also of the longitudinal parts of the correlation
tensor at ${\vec q}=0$.

The above features have to be contrasted with perturbation theory
(c.f. eq. (\ref{Apower})), where the correlation function can be
expressed by a power series in $\Gamma$ in this case and therefore
leads to a power series in $\tau$. Here the zero order term is finite
(and trivially constant in time), while the higher order terms cancel
out for each given order, c.f. (\ref{Apower}). Thus, in perturbation
theory current conservation is maintained order by order also for the
classical correlator (as one is used to from the quantum case). The
price to be payed is that in perturbation theory the infra-red
properties are completely ill.

In favor of the finite width description one should realize that {\em
any finite width calculation up to a certain order $n_x$ includes the
perturbation theory or QPA results up to the very same order $n_x$}. In
other words, if one would expand all finite width terms within the
order $n_x$ into powers of $\Gamma$ one recovers all the QPA or
perturbation theory terms up to that order!

Only in the above sense current conservation, gauge invariance and
other identities related to them (e.g. Ward identities) can be
understood and expected to hold for a finite set of terms. These
considerations are general. In particular they also apply to any
finite set of self-energy diagrams in the quantum case, which use full
Green's functions with {\em damping} as discussed in the next section.

This completes the formal discussion of the classical radiation rate
and the corresponding classical expressions for the self energy of the
photon.

\section{Non Equilibrium Green's Function Description}

In this section the production rate or photon self energy
(\ref{Npi-+}) is discussed in the context of non-equilibrium quantum
field theory, in terms of Green's functions and the corresponding
non-equilibrium diagrams. Thereby one has to go beyond perturbation
theory which is not applicable for strongly interacting systems.  The
idea to formulate theory in terms of appropriately defined physical
terms was very fruitful. It resulted in the development of the QPA
method, see \cite{LL,M}, where the changes in the real part of the
fermion self energy are taken into account substituting the free
particle energy $\epsilon^0_{\vec p}$ by the corresponding QPA energy.
This method proved to be very successful for the case of equilibrium
matter at rather small temperature ($T\ll \epsilon_F$, where
$\epsilon_F$ is the Fermi energy), since higher order corrections lead
to additional $(T/\epsilon_F)^2$ factors \cite{V1}.

Since in the QPA the imaginary part of the self energy
($\Im\Sigma_F^{R}$) is supposed to be negligible in the corresponding
Green's functions it still suffers the same infra--red problems as the
perturbation expansion.  Any finite set of diagrams leads to
infra--red divergences in the soft limit ($q \rightarrow 0$). Even
certain resummation methods as the hard thermal loop expansion
\cite{BP,BPS} in QCD do not cure the problem.  On the other hand the
classical considerations of the previous section clarify, that all
infra--red divergences disappear, if one properly accounts for the
finite collision rate $\Gamma$.  Thus, one has to avoid the zero-width
perturbation theory or QPA and seriously account for the finite
damping width $\Gamma=-2\Im\Sigma^{R}$ of the source particles.

Therefore the simplest and physically most meaningful step is to go to
the {\em full} Green's functions ${\boldsymbol G}$ (full lines in
diagrams) associated with the in-medium propagation for the
constituents of the source.  Thus one has to solve Dyson's equation in
order to include also the damping width of the particles. This is done
in sect. 4.2.  The derivation of transport schemes is summarized.
Further on some physically meaningful resummations of the diagrams are
proposed. Various graphical contributions to the self energy diagram
(\ref{grafpi}) are discussed in the QPA, and in the QC limit.

\subsection{Vertices and Green's functions}

For a theory of fermions interacting with bosons the basic
diagrammatic elements are the corresponding Green's functions $\ii G$
(lines) and interaction vertices $-\ii V$. For non equilibrium
description it makes no sense to discuss amplitudes. Rather one
evaluates the time-dependence of the ensemble averaged expectation
values of observables. Such expectation values involves standard
time-ordered products (${\boldsymbol {\hat T}}$) for the evolution of
the ''ket'' $\left|\hspace*{3mm}\right>$ and also anti--time ordered
products (${\boldsymbol {\hat A}}$) for the evolution of the ''bra''
$\left<\hspace*{3mm}\right|$ in the matrix elements. Both can be
summarized to a contour-ordered product \cite{Schwinger}.  Therefore
one distinguishes two types of vertices: the ($-$) vertices with value
$-\ii V$ pertain to the time ordered part, and the adjoint ($+$)
vertices with value $+\ii V$ for the anti--time ordered section, in
the here used convention\cite{LP}.  Thereby $V$ is the real
interaction vertex of the Lagrangian. Correspondingly two point
functions, like the Green's functions, have four components, which can
be arranged in matrix form (we reserve bold face notation for the two
by two $\{-+\}$ matrices)
\begin{equation}\label{matr}
  \ii{\boldsymbol G}_{12}:=
\left(\begin{array}{cc}\ii G^{--}_{12}&\ii G^{-+}_{12}
\\ \ii G^{+-}_{12}&\ii G^{++}_{12}\end{array}\right)
=\left(\begin{array}{lll}
\left<{\boldsymbol {\hat T}}\Psi(1)\Psi^\dagger(2)\right>&\quad&
\mp\left<\Psi^\dagger(2)\Psi(1)\right>\\[3mm]
\left<\Psi(1)\Psi^\dagger(2)\right>&&
\left<{\boldsymbol {\hat A}}\Psi(1)\Psi^\dagger(2)\right>
\end{array}\right)
\end{equation}
Here $1$ and $2$ denote the two space--time points and
$\left<\dots\right>$ the ensemble average.  Upper and lower signs
refer to fermion and boson Green's functions, respectively.  The four
Green's function components are obviously not independent. Rather they
relate to the retarded and advanced ones by
\begin{eqnarray}\label{RAgrin}
    G^R:=&-\ii\left<\left[\Psi (1),\Psi^\dagger(2)\right]_{\pm}\right>
         \Theta(t_1-t_2)&=
         G^{--}-G^{-+}=G^{+-}-G^{++}\cr
    G^A:=&+\ii\left<\left[\Psi (1),\Psi^\dagger(2)\right]_{\pm}\right>
         \Theta(t_2-t_1)&=G^{--}-G^{+-}=G^{-+}-G^{++}\;,
\end{eqnarray}
where $\Theta$ is the step function and $[\dots,\dots]_{\pm}$ denotes
the fermion anti- or boson commutator.

The unperturbed Green's function ${\boldsymbol G}^0_{12}$ is resolvent
of the corresponding free single particle Schr\"odinger - (non
rel. fermions) or Klein-Gordon equation (rel. bosons)
\begin{equation}\label{resolv}\label{SK-G}
S_1 {\boldsymbol G}^0_{12}=\left(S_2\right)^*{\boldsymbol
G}^0_{12}=\delta(1-2)
{\boldsymbol \sigma}_z\; ,\quad
S=\left\{\begin{array}{ll}\ii\partial_t+\frac{\Delta}{2m}
\quad&\mbox{(non rel. F)}\cr
-\partial_t^2+\Delta-m^2\quad&\mbox{(rel. B)}\end{array}\right.
\end{equation}
where the subscript specifies the coordinate to differentiate. Here
${\boldsymbol \sigma}_z$ is the third Pauli matrix and $\delta(1-2)$
is the four-space delta function.

With this extension to the two types of vertices $-$ and $+$ and the
corresponding four Green's function components all standard
diagrammatic Feynman rules defined for amplitudes can directly be
generalized.  These rules are then again defined with respect to the
{\em real time} (as in zero temperature theory)\footnote{Please note
that in refs. \cite{Schwinger,KB,D} the diagrammatic rules are defined
with respect to the closed time contour, which leads to the same
definition for Green's functions, however the off-diagonal components
of the self energies used there $\Sigma^{<}=-\Sigma^{-+}$ and
$\Sigma^{>}=-\Sigma^{+-}$ have opposite values.}.  For conventions and
a detailed explanation of the {\em diagrammatic rules} used here we
refer to the text book of Lifshitz and Pitaevskii \cite{LP}. For a
given $n$-point function all $n$ external vertices have a specified
sign assignment, while one has to sum over all possible sign
combinations at the internal vertices.

\subsection{Resummations: Dyson - and Kadanoff-Baym equations}
Since diagrammatic elements, e.g. Green's functions $G$ and self
energies $\Sigma$ are connected at a given vertex, the same vertex
type $-$ or $+$ appears in both functions and Dyson's equation can
simply be written in matrix form
\begin{eqnarray}\label{Dyson}
  {\boldsymbol G}_{12}&=&{\boldsymbol G}^0_{12}+ \int\di 3\di 4\;
  {\boldsymbol G}^0_{13}{\boldsymbol\Sigma}_{34} {\boldsymbol G}_{42}
   \quad\mbox{or simply as}\\[3mm]
   {\boldsymbol G}&=&{\boldsymbol G}^0+ {\boldsymbol
   G}^0\odot{\boldsymbol\Sigma}\odot {\boldsymbol G}
   ={\boldsymbol G}^0+ {\boldsymbol
   G}\odot{\boldsymbol\Sigma}\odot {\boldsymbol G}^0\nonumber
\end{eqnarray}
involving usual matrix algebra, which automatically provides the sum
over all internal vertex sign combinations.  Here the two by two
matrix ${\boldsymbol \Sigma}$ denotes the proper self energy of the
source particles\footnote{We reserve ${\boldsymbol \Sigma}$ for the
self energy of the source particles, while ${\boldsymbol \Pi}$ denotes
the self energy of the external photon.}. The $\odot$ abbreviates the
space-time folding. In diagrams Dyson's equation becomes
\begin{equation}\label{Dysondiag}
\unitlength10mm\begin{picture}(8,2.5)
   \put(0,0){\DysonB}\put(0,1.5){\DysonF}
\end{picture}\end{equation}
where we explicitly distinguish between fermions (label $F$; full
straight lines for ${\boldsymbol G}_F$) and bosons (label $B$; full
wavy lines for ${\boldsymbol G}_B$). The four components of
$-\ii{\boldsymbol \Sigma}$ are defined as the sum of all standard
proper self energy diagrams like in normal perturbation theory, now
however with definite $+$ or $-$ assignments at the external vertices,
and summed over the $-+$ signs at all internal vertices as explained
above.

Using the resolvent properties (\ref{resolv}) of ${\boldsymbol G}^0$
one can transform Dyson's equation into a set of integro-differential
equation which in short matrix notation reads
\begin{eqnarray}\label{Dysondiff}
S_1{\boldsymbol G}_{12}&=&\delta(1-2){\boldsymbol \sigma}_z
+{\boldsymbol \sigma}_z\int\di 3\;{\boldsymbol \Sigma}_{13}
{\boldsymbol G}_{32}\cr
\left(S_2\right)^*{\boldsymbol G}_{12}&=&\delta(1-2){\boldsymbol \sigma}_z
+\int\di 3\;
{\boldsymbol G}_{13}{\boldsymbol \Sigma}_{32}{\boldsymbol \sigma}_z
\end{eqnarray}
The four equations involving the time changes of $G^{-+}$ and $G^{+-}$
are known as the Kadanoff-Baym equations \cite{KB}, originally derived
by the imaginary time method. Here they are a direct consequence of
the Dyson equation in matrix form. Like the four components of
${\boldsymbol G}$ also those of ${\boldsymbol \Sigma}$ are not
independent. Their dependence can be determined observing that the
Dyson equations for retarded or advanced Green's functions have to
involve only retarded or advanced entities, respectively. Thus they
completely decouple
\begin{equation}\label{Rdys}
G^{R}=G^{0R}+G^{0R}\odot\Sigma^{R} \odot G^{R},\qquad
G^{A}=G^{0A}+G^{0A}\odot\Sigma^{A} \odot G^{A},
\end{equation}
where $\Sigma^R$ and $\Sigma^A$ are the corresponding advanced and
retarded self energies.  From (\ref{RAgrin}), (\ref{Dyson}) and
(\ref{Rdys}) one therefore follows that
\begin{eqnarray}\label{RAsig}
   \Sigma^R=\Sigma^{--}+\Sigma^{-+}\;
   ,\quad\Sigma^A=\Sigma^{--}+\Sigma^{+-}\;,\quad
   \Sigma^{++}+\Sigma^{--}=-\left(\Sigma^{+-}+\Sigma^{-+}\right).
\end{eqnarray}

The full Green's functions account for the finite damping width
\begin{equation}\label{Gamma}
\Gamma=-2\Im\Sigma^R=\ii\left(\Sigma^{-+}-\Sigma^{+-}\right)
\end{equation}
of the particles, which destroys the sharp relation between energy and
momentum. Thus the spectral function
\begin{equation}\label{spectral}
A(x;p):=-2 \Im G^R(x;p)=
\ii\left(G^{+-}(x;p) -G^{-+}(x;p)\right)
\end{equation}
is no longer an on-shell $\delta$-function by rather has a width
$\Gamma(x;p)$. This width does not only arises from decays
(resonances) but also from collisions of the particles in dense
matter.  It is important to realize, that the resummation to full
Green's functions (bold straight or wave lines in diagrams) reduces
the set of all diagrams to a subset of diagrams with {\em skeleton
topology}, where all self energy insertions are excluded in the
diagrams.

\subsection{Transport equations}

In the presented picture the Wigner transforms of the off-diagonal
Green's functions $\mp\ii G^{- +}$ and $\ii G^{+ -}$ are {\em Wigner
densities in four space and four momentum} for the {\em occupied} and
{\em available} 'single particle states', which now have a finite
width and therefore can be {\em off} mass shell.  We will see that the
corresponding non-diagonal components of the proper self energy
${-\ii\boldsymbol \Sigma}$ are the gain and loss coefficients for the
transport description of these Wigner densities.

The entry to such transport equations is given by the Kadanoff-Baym
equations, which are contained in the set
(\ref{Dysondiff}). Subtracting the two equations for $G^{-+}$ yields
\begin{eqnarray}\label{DG-+}
&&\left.
\begin{array}{r}
   \left(\ii\left(\partial_{t_1}+\partial_{t_2}\right)
+\frac{\Delta_1-\Delta_2}{2m}\right)\\
\left(-\partial_{t_1}^2+\partial_{t_2}^2
+\Delta_1-\Delta_2\right)
\end{array}\right\}G^{-+}_{12}\\ &&\hspace{1cm}=\int\di
3\left\{\Sigma^{--}_{13}G^{-+}_{32}
+\Sigma^{-+}_{13}G^{++}_{32}+
G^{--}_{13}\Sigma^{-+}_{32}+G^{-+}_{13}\Sigma^{++}_{32}\right\}\nonumber
\end{eqnarray}
for non rel. fermion or rel. bosons, respectively. The next step is to
take the Wigner transformation of this equation. This involves the Wigner
transformation of convolution integrals $C(1;2)=\int \di 3
A(1;3)B(3;2)$ which formally can be obtained through
\begin{eqnarray}
C(x;k) &=& \mbox{exp}[\frac{\ii}{2}(\partial_{k}^{A}\partial_{x}^B -
\partial_{x}^A \partial_{k}^B )] A(x;k)B(x;k)\cr
&\simeq& A(x;k)B(x;k)+\frac{\ii}{2} \{A,B\},
\end{eqnarray}
where $A(x;k)$, etc. are the Wigner transforms of $A$, $B$ and $C$
and the differential operators act on $A$ and $B$ separately. The
approximate expression in terms of 4-dimensional Poisson bracket
$\{A,B\}=\partial_kA\partial_xB-\partial_xA\partial_kB$ defines the
first order {\em gradient expansion}. To this order separating real
and imaginary parts one obtains for the Wigner densities $\mp\ii G^{-+}$
\cite{BM}
\begin{eqnarray}\label{KBeq}
   &&\left\{S(k)-\Sigma(x,k),\ii G^{-+}\right\}
-\left\{\ii G(x,k),\Sigma^{-+}(x,k)\right\}\cr
   &&=\Sigma^{-+}(x,k)G^{+-}(x,k)-\Sigma^{+-}(x,k) G^{-+}(x,k)\\
&&\mbox{with}\quad\Sigma=\left(\Sigma^R+\Sigma^A\right)/2,
             \quad G=\left(G^R+G^A\right)/2.\nonumber
\end{eqnarray}
Here $S(k)=\epsilon-{\vec k}^2/(2m)$ or $S(k)=k^2-m^2$ is the Wigner
form of the Schr\"odiger or Klein-Gordon operator (\ref{SK-G}).  The
first Poisson bracket on the left side gives the usual Vlasov part. On
the right side $\Sigma^{-+}G^{+-}$ and $\Sigma^{+-}G^{-+}$ define gain
and loss terms. This transport equation is quite general and as the
original Dyson equation still accounts for off-mass-shell processes,
part of which are contained in the peculiar second Poisson bracket
according to Botermans and Malfliet \cite{BM}. We will see that the
account of the finite damping width and the inclusion of higher order
diagrams for the self energies (sect. \ref{Decomp}) are essential to
extend the scheme beyond classical transport concepts.

Equation (\ref{KBeq}) by itself is not yet complete, since it requires
a relation a relation between $G^{-+}$ and $G^{+-}$, which normally is
provided by a certain physical ''ansatz''.  Under near on-shell
conditions, where the Wigner densities are well peaked close to a
single on-shell energy $\epsilon\sim\epsilon_{\vec p}$, c.f. next
subsection, one can use the Kadanoff - Baym ansatz or apply the QPA
(see Appendix A). Then the integral over $\epsilon$ reduces
(\ref{KBeq}) to a familiar form \cite{D,VBRS}

\begin{eqnarray}\label{KBeqsimple}
\left\{\partial_t +\vec{v}\partial_{\vec{x}}  -
\left(\partial_{\vec{x}} \epsilon_{\vec{p}}\right)\partial_{\vec{p}}
\vphantom{\int}\right\}
&& n({\vec x},{\vec{p}},t)\cr
=&&\left( \pm \ii \Sigma^{-+}(1\mp n({\vec x},{\vec{p}},t))
    +\ii\Sigma^{+-} n({\vec x},{\vec{p}},t)\right) Z_{\vec{p}}\\
=&&\left(\pm \ii \Sigma^{-+}-\Gamma n({\vec x},{\vec{p}},t)\right)
Z_{\vec{p}} \nonumber ,
\end{eqnarray}
for the on-shell particle densities of non-rel. fermions or rel. bosons
\begin{equation}\label{part-dens}
n_F({\vec x},{\vec{p}},t)= -\ii \int \frac{d\epsilon}{2\pi}G^{-+}_F\;, \quad
n_B({\vec x},{\vec{p}},t)= \ii\int \frac{d\epsilon}{2\pi}2\epsilon G^{-+}_B .
\end{equation}
Here $\vec{v}=\partial_{\vec{p}} \epsilon({\vec{p}})$ is the group
velocity, and $Z_{\vec{p}}=1/(1 - {\partial
\mbox{Re}\Sigma^R}/{\partial \epsilon})$ for non-relativistic fermions
and $Z_{\vec{p}}= 1/(2\epsilon - {\partial
\mbox{Re}\Sigma^R}/{\partial \epsilon})$ for relativistic bosons is a
normalization factor. In all expressions $\epsilon$ is determined by
the dispersion relation $\epsilon =\epsilon({\vec{p}})$,
c.f. (\ref{ep}) below.

The form of the collisional integral in (\ref{KBeqsimple}) is
convenient in many cases, e.g.  to extract stationary solution
(cf. next subsection) or for the relaxation time approximation, where
$1/\Gamma Z_{\vec{p}}$ is the corresponding collision time, see
(\ref{taun}). In the diffusion approximation of eq.(\ref{KBeqsimple})
one easily recovers eq.(\ref{FP}). The usual Boltzmann-like form of
the collisional integral is obtained from the lowest order self energy
diagram which in QPA contributes to $\Im \Sigma^R$ (usually second
order in the two body interaction), c.f. \cite{D,VBRS} and
(\ref{QPAself}) in sect. \ref{QPdecomp}. More general schemes beyond
QPA are suggested in sect. \ref{keydiag}. The photon production rate,
eq. (\ref{dN}), is a simple case of (\ref{KBeqsimple}).

\subsection{Stationary and Equilibrium Properties}
For a homogeneous stationary system in Wigner $(x;p)$ representation
all Green's functions become space-time independent and the $\odot$
operation reduces to a simple pro\-duct for the remaining 4-momentum
part in the Dyson equation (\ref{Dyson}). With $p=(\epsilon,{\vec p})$
we therefore drop the $x$-argument.  Eqs. (\ref{Rdys}) can then be
solved algebraically
\begin{equation}\label{G^Req}
   G_F^R(p)=\frac{1}{\epsilon+\mu_F -\epsilon_{\vec p}^0-\Sigma^R(p)}\; ,
  \quad  G_B^R(p)=\frac{1}{(\epsilon+\mu_B )^2-(\epsilon_{\vec p}^0)^2
   -\Sigma^R(p)}\; ,
\end{equation}
while $G^A(p)=(G^R(p))^*$ and $\Sigma^A(p)=(\Sigma^R(p))^*$.
The dispersion relations are
\begin{equation}
\label{ep}\begin{array}{rlrll}
\epsilon+\mu_F =& \epsilon_{\vec p}^0+\Sigma^{R}_F (\epsilon+
\mu_F ,
\vec{p}),&\quad\quad\epsilon_{\vec p}^0=&
\vec{p}^2/2m\quad&\mbox{(non-rel. fermions)}\cr
(\epsilon+\mu_B )^2=&(\epsilon_{\vec p}^0)^2+\Sigma^{R}_B
(\epsilon+\mu_B ,\vec{p}),&\quad(\epsilon_{\vec p}^0)^2
=&\vec{p}^2+m^2 \quad&\mbox{(rel. bosons)}\end{array}
\end{equation}
with corresponding free on-shell energies $\epsilon_{\vec p}^0$ and
chemical potentials $\mu_B$ and $\mu_F$. With
$ \Gamma(p)=-2\Im \Sigma^R(p)=
   \ii\left(\Sigma^{-+}(p)-\Sigma^{+-}(p)\right)$
the spectral function (\ref{spectral}) are
\begin{eqnarray}
   A_F(p) &=& \frac{\Gamma(p)}{\left(\epsilon+\mu_F -\epsilon_{\vec
   p}^0-\Re\Sigma^R(p)\right)^2+(\Gamma(p)/2)^2}
   \quad\mbox {\rm (non-rel. fermions)},\cr
   A_B(p)&=&\frac{\Gamma(p)}{\left((\epsilon+\mu_B )^2
  -(\epsilon_{\vec p}^0)^2
  -\Re\Sigma^R(p)\right)^2+(\Gamma(p)/2)^2}
   \quad\mbox{\rm (rel. bosons)}.
\end{eqnarray}
They satisfy the sum rules
\begin{equation}\label{SRA}
\int_{-\infty}^{\infty} A_B (p)2\epsilon \frac{\di\epsilon}{2\pi}=1
\quad\mbox{and}
\int_{-\infty}^{\infty} A_F (p) \frac{\di\epsilon}{2\pi}=1.
\end{equation}
The generalization to relativistic fermions with gamma-matrices up to
tensor interactions can be found in ref. \cite{PHenning2}. For
illustration and later use we give the equilibrium results explicitly,
which follow from the stationary condition
$\Sigma^{-+}(p)G^{+-}(p)=\Sigma^{+-}(p) G^{-+}(p)$,
c.f. eq. (\ref{KBeq}), and the
Kubo-Martin-Schwinger condition \cite{KMS}
\begin{equation}
\Sigma^{-+}(p)=\Sigma^{+-}(p)e^{-\epsilon/T}.
\end{equation}
Then all the Green's
functions can be expressed through either retarded or advanced Green's
functions.  From (\ref{matr}), (\ref{RAgrin}) and (\ref{RAsig}) one then
finds
\begin{eqnarray}\label{Geq}
      G^{--}(p)&=(1\mp n_\epsilon)G^R(p)\pm n_\epsilon G^A(p)\; ,\quad
    &G^{-+}(p)=\pm \ii n_\epsilon A(p),\\
    G^{+-}(p)&=-\ii(1\mp n_\epsilon)A(p)\; ,\quad
    &G^{++}(p)=-(1\mp n_\epsilon)G^A(p)\mp n_\epsilon G^R(p)\; ,
\nonumber
\end{eqnarray}
and the relations for the four components of the self energies
\begin{eqnarray}
   \Sigma^{--}=&\Sigma^R\pm \ii n_\epsilon\Gamma(p)\; ,
   \quad
   &\Sigma^{-+}=\mp \ii n_\epsilon \Gamma(p),\cr
   \Sigma^{+-}=&\ii(1\mp n_\epsilon)\Gamma(p)\; ,
   \quad&\Sigma^{++}=-\left(\Sigma^{--}\right)^*\; ,
\end{eqnarray}
where the thermal occupations at temperature $T$ (Fermi-Dirac or
Bose-Einstein distributions) are
\begin{equation}
n_\epsilon =\{\exp[\epsilon/T]\pm 1\}^{-1}\; .
\end{equation}

In the general non--equilibrium case there are no such simple relations
between Green's functions and self energies as at equilibrium. In order
to proceed one may simplify the problem applying so called
Kadanoff--Baym ansatz (given in Appendix A) or using the QPA.

\subsection{Diagrammatic Decomposition into Physical
Sub-Processes }\label{Decomp}

There have been many attempts in the literature to eliminate the
redundancy in the definition of the four Green's function components.
We do not like to follow such schemes and rather prefer to keep all
four components as they are, since they display a symmetry between the
time-ordered and the anti-time ordered parts, i.e. between the "bra" and
"ket" parts of the correlator\footnote{In thermal field theory the ''$+$''
vertices are often considered as ghosts. We do not like to support
this viewpoint as these conjugate vertices are as physical as the ''$-$''
ones for the full correlation matrix element!}.

In this formulation physical observables like densities, production
rates, etc., are always given by diagrams, where the external vertices
appear in conjugate pairs, i.e.  with fixed opposite signs, just
defining correlation functions. If Fourier transformed over space-time
differences they have the properties of Wigner functions. Special
examples are the off-diagonal components of the Green's functions and
self energies, which define Wigner densities and gain or loss rates,
respectively.  Working with full Green's functions such correlation
diagrams are given by a sum of all topologically distinguished
skeleton diagrams with the fixed external vertices of opposite
signs. Each diagram of given topology consists of $2^\nu$ terms, where
$\nu$ is the number of internal vertices, due to the $-+$
summations. These extra summations make this approach rather
non-transparent. In this section, however, we like to suggest a very
simple classification of correlation diagrams and a reformulation of
the corresponding sum, which amends a simple physical interpretation.

We start with the observation, that at least in one way any
correlation diagram of given topology and given sign assignments at
all vertices can be decomposed into two pieces, such that
each of the two sub pieces is a connected diagram which carries only
one type of sign on all its external vertices
\footnote{For the construction: just deform the diagram such that all
$+$ and $-$ vertices are placed left and respectively right from a
vertical division line and cut along this line. Pieces which then
become disconnected are to be reconnected to the other side until two
connected sub-pieces remain. In case that disconnected pieces appear on
both sides the result may depend on the order of reconnection and
consequently different decompositions are possible.  }
\begin{equation}\label{decomp}\unitlength8mm
\begin{picture}(12.0,1)
\put(0,.02){\decomposition}
\end{picture}
\end{equation}

The reason for such a decomposition is that then each diagram is given
by a "product" of two sub diagrams $\alpha^-$ and $\beta^+$ with the
same external lines, which one may call amplitude -- and adjoint
amplitude diagrams. Here $\alpha$ and $\beta$ denote amplitude
diagrams including sign assignments. The adjoint $\alpha^+$ of any
amplitude diagram $\alpha^-$ is given by inverting the senses of all
propagator lines and inverting the signs of all vertices; the
respective values are conjugate complex to each other
$(\alpha^-)^*=\alpha^+$. For amplitude diagrams only the external
vertices have definite signs, while internal vertices have still no
sign restrictions. In analytical terms the decomposition
(\ref{decomp}) can be written as
\begin{eqnarray}\unitlength6mm
\begin{picture}(4.,1)
\put(-1.3,.2){\decomposition}
\end{picture}
&=&(\mp\ii) G^{-+}(p_1)\dots \ii
   G^{+-}(p_m)\beta^+(p_1,\dots,p_m)\alpha^-(p_1,\dots,p_m)\\
   \hspace*{3cm}&=&(\mp\ii) G^{-+}(p_1)\dots \ii G^{+-}(p_m)
   \left(\beta^-(p_1,\dots,p_m)\right)^*\alpha^-(p_1,\dots,p_m)
   \nonumber
\end{eqnarray}
for space-time homogeneous cases (in general corresponding space-time
foldings appear).  Here $\mp\ii G^{-+}(p_k)$ or $\ii G^{+-}(p_k)$
(depending on the line-sense) are the Wigner densities of occupied (or
available states) for each external line, with off-shell 4-momenta
$p_1$ to $p_m$, connecting both pieces. When a correlation diagram can
be decomposed in more than one way according to rule (\ref{decomp}),
the picture is not unique and one may assign partial weights which sum
to unity to the different decompositions\footnote{ A simple example
for such an ambiguity is the following diagram which permits two
decompositions indicated by the thin line:\\[-2mm]\hspace*{8cm}
\unitlength6mm \twodecomp}. Applying this decomposition scheme to all
skeleton diagrams of the proper self energy leads to a decomposition
in terms of physical processes with a varying number $m$ of external
''states'' $p_1,\dots ,p_m$ besides the external photon in this case
\begin{equation}
   -\ii\Pi^{-+}(q)=\sum_{m=2}^\infty\int
   R^{-+}(p_1,\dots,p_m)\ii G^{-+}(p_1)\di ^4p_1\dots
  \ii G^{+-}(p_m)\di^4p_m.
\end{equation}
Here the $R^{-+}(p_1,\dots,p_m)$ define the partial rates for each
physical sub process with certain {\em in}-- and {\em out}-states
$p_1$ to $p_m$, which all can be {\em off-shell}. In the sense of
crossing symmetry the notion of {\em in} and {\em out} may be fixed by
the line sense: incoming or outgoing lines. Naturally for fermions
there are as many in-- as out-states.  It is important to realize that
the partial rates
\begin{equation}\label{Rp1-pm}
   R^{-+}(p_1,\dots,p_m)={\sum_{\alpha,\beta}}^\prime
   \left(\beta^-(p_1,\dots,p_m)\right)^*\alpha^-(p_1,\dots,p_m)
\end{equation}
arise from a {\em restricted} sum (indicated by the prime at the
sum-symbol) over amplitude products $\alpha$ times $\beta^*$ such that
each term arises from a given correlation diagram in the original
sum. Note in particular that the sum (\ref{Rp1-pm}) {\em no longer}
includes all possible interference terms of any two amplitudes!  The
unrestricted sum over all pair-products of amplitude $\alpha^-$ times
$(\beta^-)^*$ is false and leads to serious inconsistencies, since
there are interference terms, which correspond to closed diagrams of
{\em non-skeleton type, which have to be omitted!} The
simplest case to see this is the normal diagram for bremsstrahlung
(left)
\begin{equation}\label{nonskeleton}
\unitlength8mm
\begin{picture}(10,1.5)
     \put(0,-.8){\nonskeleton}\end{picture}.
\end{equation}
Its absolute square leads to a diagram with self-energy insertion
(\ref{nonskeleton};right), which is not a proper skeleton diagram;
rather the corresponding partial rate is already included in the
$1$-loop diagram with full Green's functions!

Therefore, only when each term that enters a physical rate
(\ref{Rp1-pm}) originates from a valid decomposition of proper
skeleton self energy diagrams, then this formulation is void of double
counting. It amends a straight forward interpretation in terms of
physical scattering processes between certain in -- and
out-states. These processes occur with partial rates $R$, which can be
positive or negative. The warning to be formulated at this instance
is, that if one works with full Green's functions which also include
the finite damping width $\Gamma$, particular care has to be taken.
The unconsidered account of certain Feynman amplitudes together with
finite width spectral functions for the particles, as sometimes done
heuristically, may lead to serious inconsistencies.

On the other hand the important point to realize at this level is that
the so defined decomposition gives rise to a generalized formulation
of transport theories, where multi-particle processes can be
considered in a well defined scheme even if one permits for off--shell
propagation.  Note in particular: for a theory of fermions interacting
with bosons the contribution with the fewest number of external
particles is just three (rather than four as in the Boltzmann
equation). It results from the decomposition a one-loop diagram and
allows for one particle $in$ and two $out$ and vice versa, . Thus, in
dense matter an off-shell fermion can just decay into a fermion plus
boson or the opposite can happen, e.g. see \cite{VBRS,V4}.  For these
processes it is important that all particles have a finite damping
width in dense matter, so that creation and decay modes, which are
forbidden from energy-momentum conservation in free space, may occur
without principle restrictions in dense matter.

Please note that our decomposition rules are different from the
standard cutting rules which only apply either to the set of
perturbation theory diagrams \cite{cuttingrules} or to the set of the
quasi-particle diagrams \cite{V1}. There diagrams are cut across all
$+-$ lines, each cut providing an on-shell delta function from the
zero width spectral functions. Diagrams that can be cut into more than
two pieces contain a product of more than one delta-function on total
energy conservation and therefore show a singular behavior which is
not present in the final correlation function. Therefore all such
"multi"-cut diagrams have to cancel out. Such arguments can no longer
be given in the case that all spectral functions have finite widths,
where the set of diagrams is reduced to the set of skeleton diagrams.
Thus, such diagrams have also to be considered in a description with
full Green's functions!

\subsection{Three and Four Point Functions}

In principle one can stick to the above picture since all infra-red
divergences disappear for all diagrams due to the finite propagation
times of all Green's functions. However one may have to consider still
quite a numerous amount of diagrams in order to achieve meaningful
results.  For instance on the QP level we expect, that the
proper {\em in-medium} current appears at both external vertices in a
symmetric fashion. Thus for the convective currents one expects
\begin{eqnarray}\label {classicalcurrent}
j^{\nu}&\simeq&ev^{\nu}=e\frac{\partial}{\partial p_{\nu}}\epsilon({\vec p})
   =e\left(\frac{\partial}{\partial p_{\nu}}\epsilon^0_{\vec p}
     +\frac{\partial}{\partial p_{\nu}}\Re\Sigma^R \right)/
      \left(1-\frac{\partial}{\partial \epsilon}\Re\Sigma^R \right),
\end{eqnarray}
where $\epsilon ({\vec p})=\epsilon^0_{\vec p}+\Re\Sigma^R
(\epsilon({\vec p}),{\vec p})$ defines the QP energy momentum
relation, see eq. (\ref{eps}) in Appendix A. In diagrammatic terms
this can be achieved by certain partial resummation that lead to
vertex corrections. Thereby we shall not consider an immediate
resummation to the exact full vertex, since this would amount to solve
the whole problem. Rather we like to stay to a picture where in
certain limits like the QPA and QC limit, piece by piece an
interpretation in physically meaningful terms can be given.

The considerations in the preceding subsection assigned a particular
role to the full $-+$ and $+-$ Green's functions as Wigner
densities. This suggests to apply further resummations and to extend
the ideas put forward in ref. \cite {V1} in the context of
quasi-particles now to particles with finite width.  Namely, one likes
to gather diagram pieces that are void of the Wigner densities
$G^{-+}$ or $G^{+-}$, both for fermions and bosons. That is, one likes
to resum sub-pieces of skeleton diagrams with given number and type of
external vertices (3 point or 4 point functions, for example) where
all internal and external vertices have only one definite sign
value. The $\{-\}$ diagrams then contain only time-ordered full
Green's functions $G^{--}$ and therefore represent a straight forward
generalization of the standard zero temperature Feynman 3 or 4 point
functions now including the full self energies.  The $\{+\}$ diagrams
are just the adjoined expressions. This way one can define 4 point
functions (in-medium interactions)
\begin{equation}\label{4point}
   \unitlength10mm\begin{picture}(14,1.2)\put(0,-.6){\fourpointeq}
   \end{picture}
\end{equation}
Here the thick wavy lines relate to the corresponding
$G_B^{--}$--exact boson propagators or two-body potentials in
non-relativistic theories with potentials. Since only like sign
vertices are permitted, no $G^{-+}$ and $G^{+-}$ lines appear in these
functions.  Such resummed expressions have been proven useful in the
low temperature QPA to define in-medium interactions and
effective vertices and they appear also quite meaningful in the limit
of low densities, as in the classical limit for example.  Various
approximation levels are possible for the 4-point functions; a
detailed discussion would be beyond the scope of this presentation. We
mention just a few possibilities:
\begin{itemize}
\item[a)] a ladder summation in the $s$-channel (horizontal in
diagram (\ref{4point})) for the particle-particle ($p$-$p$) and
particle-hole ($p$-$h$) channels generalizes the Bruckner $G$-matrix
to non--equilibrium;
\item[b)] in many practical cases (e.g.  Landau - Migdal's
Fermi-liquid theory \cite{LL,M}) the 4-point functions are
approximated by 2-point approximants in the $t$-channel (vertical in
(\ref{4point})). Then RPA-type resummations are possible
\cite{V1,MSTV}, which iterate $p$-$h$ "-- --" and "+ +" loops in the
$t$-channel; details are given in Appendix B; where also the scheme of
the bosonization of the interaction is presented,
cf. \cite{MSTV,Voskr}.
\item[d)] the ultimate could be a crossing and exchange symmetric
form; in this case one relies on suitable parameterizations.
\end{itemize}

These like--sign effective interactions generalize the two-body
scattering matrix in matter to non--equilibrium. Thereby one does not
only account for the change of the fermionic occupations (as already
considered in the literature) but also includes the damping of the
fermions.  In this respect it would be interesting to see, how bound
states (e.g. the deuteron \cite{SRS} or the $J/\Psi$) change their
properties in dense matter (Mott transition) also due to the damping
widths. For consistency these like--sign effective interactions then
also enter the definition of the in-medium vertices (3 point
functions) defined as
\begin{equation}\label{3point}
   \unitlength10mm\fullvertexeq
\end{equation}

\subsection{Key diagrams}\label{keydiag}

For simplicity we confine the discussion to the case where the
'external' photon just couples to fermions. Any generalization to
other types of particles (external and source internal) is straight
forward. Due to the above considerations from now on the remaining
diagrams include the following elements: full fermion Green's
functions, like-sign 4-point interactions and the corresponding
vertices.  Please notice that this reduces the set of diagrams even
further! In particular not all sign combinations are permitted any
longer since some of them are already included in the resummed 3 or 4
point functions.

All photon self-energy diagrams can be build up by iterative four
point insertions. Thus, the set of diagrams for $\Pi^{-+}$ reduces to
\begin{equation}\label{keydiagrams}
   \unitlength6mm\keydiagrams\; .
\end{equation}
This set of ''key''- diagrams is important for all subsequent
considerations and therefore deserves further comments.
\begin{enumerate}
\item Each diagram in (\ref{keydiagrams}) represents already a whole
class of perturbative diagrams of any order in the interaction
strength and in the number of loops. The most essential term is the
one-loop diagram\footnote{In perturbation theory or QPA the
corresponding one-loop diagram usually vanishes for on-shell photons
due to conservation laws. Here however, with full Green's and vertex
functions it represents a series of perturbative diagrams as the
reader can easily imagine.}, which is positive definite, and
corresponds to the first term of the classical Langevin result for
$\Pi_{\mathrm{cl}}$ in (\ref{Mtau}) as we shall show later. The other
diagrams represent interference terms due to rescattering.
\item Compared to conventional diagrams, vertex corrections can appear
on both sides of one loop as they are separated by $\{+-\}$ lines (see
example given in Appendix B).
\item Note that the restriction to like sign vertices for the
resummations (\ref{3point}) and (\ref{4point}) are defined with
respect to skeleton diagrams in terms of full Green's functions.
''Opened'' to perturbative diagrams with thin $G^0$ lines, these can
still contain alternative signs, since Dyson's equation (\ref{Dyson})
includes all signs in the intermediate summations!
\item In some simplified representations (being often used) the
4-point functions behave like intermediate bosons (e.g. phonons), c.f.
Appendix B.
\item For particle propagation in an external field, e.g. infinitely
heavy scattering centers, only the one-loop diagram remains, since the
one deals with a genuine one-body problem. However, extra
complications arise, since translation invariance is generally broken
and the Green's functions then also depend on $x$.
\end{enumerate}

\subsection{Decomposition of Closed Diagrams into Feynman
Amplitudes in the QPA}\label{QPdecomp}

The QPA is quite commonly used concept originally derived for Fermi
liquids at low temperatures (Landau-Migdal, see \cite{LL,M}). There
one deals with on-mass-shell fermions in matter (quasi-particles)
described by the pole part of the Green's functions, i.e. one assumes
that $\mbox{Im}\Sigma_{F}^{R} \rightarrow 0$ in the Green's function
$G_{F}^{R}$. Then with the help of some phenomenologically introduced
interaction (particle-hole irreducible) one calculates the values
$\mbox{Re}\Sigma^R$ and $\mbox{Im}\Sigma^R$ which now depend on
quasi-particle properties. Since in QPA the finite width contributions
have to appear in higher order through corresponding
$\Im\Sigma$-insertions the whole set of QPA diagrams defining the full
$-\ii\Pi^{-+}$ is by far larger, than set (\ref{keydiagrams}).

The QPA has considerable computational advantages as Wigner densities
("-- +" and "+ --" lines) become energy $\delta$--functions, and the
particle occupations can be considered to depend on momentum only
rather than on the energy variable\footnote{The later approximation is
also often used beyond the scope of the QPA and is then known as
Kadanoff--Baym ansatz, see \cite{KB} and Appendix A,
c.f. \cite{SL}}. Formally the energy integrals in eq.(\ref{grafpi}),
(\ref{keydiagrams}) can be eliminated, in diagrammatic terms just
cutting the corresponding "-- +" and "+ --" lines \cite{V1}.  This way
one establishes a correspondence between correlation diagrams
(\ref{keydiagrams}) and usual Feynman amplitudes in terms of QPA
asymptotic states and QPA Green's functions. Thus the QPA allows a
transparent interpretation of correlation diagrams.

For the dynamics of the fermion transport the first QPA diagram that
contributes to the gain (loss) term
\begin{eqnarray}\label{QPAself}
&&-\int\frac{\di\epsilon_1}{2\pi}\Sigma^{-+}(p_1)G^{+-}(p_1)
\simeq\left\{\unitlength8mm\selfinsF\right\}_{\mbox{QPA}}
(1-n_1)\\&&\longrightarrow\int\di^4p_2\dots\di^4p_4
\left|\unitlength8mm\boltzm\right|^2\delta^4(p_1+p_2-p_3-p_4)
(1-n_1)(1-n_2)n_3n_4\nonumber
\end{eqnarray}
leads to the standard Boltzmann collision term with corresponding
occupation and Pauli suppression factors for the in and out states.
Here and below the full blocks denote the effective two--fermion
interactions, and thick fermion lines denote the QPA states or Green's
functions.

For the here studied photon rates we discuss in detail the
correlations diagrams on the right side of (\ref{keydiagrams}) with
consecutive numbers 1 to 6. Thereby diagrams 1, 2, 4 and 5 describe
the bremsstrahlung related to a single in-medium scattering of two
fermionic quasi-particles and can be symbolically expressed as Feynman
amplitude (\ref{Feynman}a)

\begin{eqnarray} \label{Feynman}
   (a)&\hspace*{5mm}\unitlength8mm\begin{picture}(2.5,0.7)
   \put(0,-.35){\borndiag}\end{picture}\hspace*{4.7cm}&(b)
  \hspace*{5mm}\unitlength8mm\begin{picture}(2.5,0.7)
   \put(0,.2){\selfinsert}\end{picture}\\[3mm]
  (c)&\hspace*{5mm}\unitlength8mm\begin{picture}(4.5,1)
   \put(0,-.35){\twocol}\end{picture}\hspace*{3cm}&(d)
  \hspace*{5mm}\unitlength6mm\begin{picture}(4.5,1)
   \put(0,-.55){\intrad}\end{picture}\nonumber
\end{eqnarray}
The full circle denotes the effective vertex. One should bear in mind
that the photon may couple to any of the external fermion legs and all
exchange combinations are possible. The one-loop diagram in
(\ref{keydiagrams}) is particular, since its QP approximant vanishes
for real or time-like photons. However the full one-loop includes QPA
graphs of the type (\ref{Feynman}b), which survive to the same order
in $\Gamma/{\bar{\epsilon}}$ as the other diagrams \cite{V1}. In fact
it is positive definite and corresponds to the absolute square of the
amplitude (\ref{Feynman}a)), c.f. (\ref{nonskeleton}).  The other
diagrams 2, 4 and 5 of (\ref{keydiagrams}) describe the interference
of amplitude (\ref{Feynman}a) either with those where the photon
couples to another leg or with one of the exchange diagrams. Thereby
for neutral interactions diagram (\ref{keydiagrams}:2) is more
important than diagram 4 , while this behavior reverses for charge
exchange interactions (the latter is also important for gluon
radiation from quarks in QCD transport due to color exchange
interactions).  Diagrams like 3 describe the interference terms due to
further rescatterings of the source fermion with others.  According to
our rules the diagram (\ref{keydiagrams}:3) corresponds to a two-body
collision process and describes the interference of amplitude
(\ref{Feynman}c) with that one where the photon couples to the initial
leg.  Diagram (\ref{keydiagrams}:6) describes the photon production
from intermediate states and is given by Feynman graph
(\ref{Feynman}d). In the soft photons limit ($\omega_q\ll \epsilon_F$)
this diagram (\ref{Feynman}d) gives a smaller contribution to the
photon production rate than the diagram (\ref{Feynman}a) in QPA, where
the normal bremsstrahlung contribution diverges like $1/\omega_q$
compared to the $1/\epsilon_F$--value typical for the coupling to
intermediate fermion lines \cite{V3}. However in some specific cases
the process (\ref{Feynman}d) might be very important even in the soft
limit.  This is indeed the case for so called modified URCA process
$nn \rightarrow npe\bar{\nu}$ which is of prime importance in the
problem of neutrino radiation from the dense neutron star interior,
see \cite{V4}.

Some of the diagrams, (c.f. the graph shown in footnote 7) which are not
presented explicitly in eq.(\ref{keydiagrams}) give more than two
pieces, if being cut, so they do not reduce to the Feynman amplitudes.

For the validity of the QPA one normally assumes that
$\Gamma\ll\bar{\epsilon}$, where $\bar{\epsilon}$ is an average
particle kinetic energy ($\sim T$ for equilibrium matter). With
$\Gamma \sim \pi^2T^2/\epsilon_F$ for Fermi liquids,
c.f. (\ref{sigret}), the QPA constitutes a consistent scheme for all
thermal excitations $\Delta \epsilon \sim T\ll\epsilon_F$. However
with the application of transport models to higher energies this
concept has been taken over to a regime where its validity can no
longer easily be justified.  Moreover, our considerations show that
the condition $\Gamma\ll\bar{\epsilon}$ is not at all
sufficient. Rather one has to demand that also $\omega\gg \Gamma$ in
the QPA, since finally energy differences of order $\omega$
appear\footnote{This statement is particular, since one compares the
photon energy $\omega$ with the damping width of the source particles
$\Gamma$, while the damping rate of the photon itself
$\gamma\simeq\left|\Pi^{-+}\right|/(2\omega n^B_{\omega}) <4\pi
e^2\rho_0/(\Gamma m)\sim 20\mbox{MeV}^2/\Gamma $ for nuclear matter
can be quite small!}. In particular, the remaining series of
QPA-diagrams is {\em no longer convergent} unless $\omega>\Gamma$,
since arbitrary powers in $\Gamma/\omega$ appear, and there is no hope
to ever recover a reliable result by a finite number of QPA-diagrams
for the production of soft quanta! With {\em full Green's functions},
however, one obtains a description that uniformly covers both the soft
($\omega\ll\Gamma$) and the hard ($\omega\gg\Gamma$) regime.

\section{The Quasi-classical Limit (QC)}
In this section we like to discuss, which class of diagrams remains in the
quasi-classical limit and how this is to be interpreted.

The QC limit requires that\\[3mm]
\hspace*{.5cm}\parbox{14cm}{
\begin{itemize}
\item[$i)$] all occupations of the source
  particles are small ($\left<n_{\vec p}\right> \ll 1$) implying a
  Boltzmann gas with $\mu\ll-T$ and that
\item[$ii)$] all inverse length or time-scales times $\hbar$ are
  small compared to the typical momentum and energy scales of the source
  systems.
\end{itemize} }\\[3mm]
In particular this implies $\hbar \omega,\hbar|{\vec
q}|\ll\bar{\epsilon}$, and a collision rate $
\Gamma=\hbar/\tau_{\mathrm{coll}} \ll\bar{\epsilon}$, where
$\bar{\epsilon}$ is a typical particle kinetic energy ($\sim T$ for
equilibrium matter). To be precise: $\omega$ and ${\vec q}$ of the
produced particle are sensitive to the space-time structure of the
source, while they are negligible as far as energy and momentum
balances are concerned. The latter fact permits to prove the
Kadanoff--Baym ansatz in this case (see Appendix~A and discussion of
eq. (\ref{Imsimp}) below) which considers the occupations of the
source particles to dependent only on momentum $n_{\vec
p}=n_{\epsilon_{\vec p}-\mu_F}$ but no longer on energy $\epsilon$.
Also we assume that $\Gamma$ will not depend on time in between
subsequent collisions ($|\tau|\sim 1/\Gamma$).

We note in particular that for bosons with chemical potential $\mu_B=0
$, like the produced photon, the {\em equilibrium} occupations will be
large, $n_{B}\approx T/\omega\gg 1$!  This fact is of no further
relevance, if one excludes internal photon lines in the proper
correlation functions (\ref{Npi-+}), $\Pi^{-+}$ and $\Pi^{+-}$ as we
do.

\subsection{Time Structure of Green's Functions and Loops in QC Limit}

For fermion Green's functions one has the following simplifications
\begin{equation}\label{free}
G^{- +}_F(p) \simeq  \ii n_{\vec p} A(p),\quad
G^{+ -}_F(p) \simeq -\ii (1- n_{\vec p})A(p),
\end{equation}
while at large temperatures $T$ the particle occupations
are given by
\begin{equation}
n_{\vec p} \simeq  \mbox{exp}[-(\epsilon_{\vec p}-\mu_F)/T]\ll 1.
\end{equation}

The correspondence between the diagrammatic expansion
(\ref{keydiagrams}) and classical limit of sect. 3 becomes more
transparent if one uses the mixed $\tau-{\vec p}$ representation for
the Green's functions, where $\tau$ is the time difference between the
two space-time points. Then from the definition of $G^{- +}$ and
$G^{+-}$ Green's functions (see eq. (\ref{matr})) one immediately
finds for fermions
\begin{equation}\label{tp}
   G_F^{- +}(\tau,{\vec p}) \simeq \ii n_{\vec p} \exp [-|\Gamma\tau|/2
   -\ii\epsilon_{\vec p} \tau],
\end{equation}
\begin{equation}\label{tp2}
  G_F^{+ -}(\tau,{\vec p}) \simeq -\ii (1-n_{\vec p})\exp [-|\Gamma\tau|/2
  -\ii\epsilon_{\vec p} \tau],
\end{equation}
while $G_F^{--}=(1-n_{\vec{p}})G_F^R-n_{\vec{p}}G_F^A$ and
$G_F^{++}=-(1-n_{\vec{p}})G_F^A+n_{\vec{p}}G_F^R$ are essentially
retarded and advanced, respectively.

A further simplification comes from the time behaviour of
fermion-loops, c.f (\ref{open}), which mediate classical energy and
momentum transfers.  The corresponding time scales $1/\epsilon$ or
$1/({\vec v}{\vec p})$ are very short on the damping scale $1/\Gamma$,
so that such loop insertions become instantaneous. This is the reason,
why one recovers a Markovian description for the motion of the source
in the classical limit.  With these simplifications we now calculate
the diagrams (\ref{keydiagrams}).

\subsection{Self Energy Diagrams in the QC Approximation}

In the mixed $\tau-{\vec q}$ representation the one loop diagram is
given by
\begin{eqnarray}\label{onelooptau}
-\ii\Pi^{-+}_{0}(\tau,{\vec q})&=&\int\frac{\di^3p_1\di^3p_2}{(2\pi)^6}
V^\mu V^\nu
G^{-+}(\tau,{\vec p}_1)G^{+-}(\tau,{\vec p}_2)(2\pi)^3
\delta({\vec p}_2-{\vec q}-{\vec p}_1)\cr
&\simeq&\int\frac{\di^3p}{(2\pi)^3}
 V^\mu V^\nu n({\vec p}+{\vec q}/2)(1-n({\vec p}-{\vec q}/2))
 e^{-|\Gamma\tau|}e^{-\ii({\vec q}{\vec v})\tau},
\end{eqnarray}
if $|{\vec q}{\vec v}|\ll T$. Here $V^\mu\approx\sqrt{4\pi}j^{\mu}
({\vec p})$ defines the in-medium photon - fermion vertex in the
classical limit following eq. (\ref{classicalcurrent}), while $\vec
p\approx(\vec p_1 +\vec p_2)/2$. Apart from the ${\vec q}$-dependent
oscillations the time structure of this diagram is given by an
exponential decay: $e^{-|\Gamma\tau|}$ which leads to
\begin{eqnarray}\label{oneloopomega}
-\ii\Pi^{-+}_0(\omega,{\vec q})&=4\pi e^2&\int\frac{\di^3p}
{(2\pi)^3} n({\vec p}+{\vec q}/2)(1-n({\vec p}-{\vec q}/2))
\frac{2\Gamma v^iv^k}{\left(\omega-{\vec
q}{\vec v})
\right)^2+\Gamma^2}\cr
&\approx&4\pi e^2\rho_0\left<\frac{2\Gamma v^iv^k}{\left(\omega-{\vec
q}{\vec v})
\right)^2+\Gamma^2}\right>\; .
\end{eqnarray}
for the spatial components of $\Pi^{-+}$.  This expression is
identical to the $n=0$ term of the classical Langevin result
(\ref{Mtaun0}).

The classical Langevin example (sect. 3) considers the propagation of
a single charge (say a proton) in neutral matter (e.g
neutrons). Therefore for this case only diagrams occur, where both
photon vertices attach to the same proton line. Also, of course, no
direct proton-proton interactions occur.  In the following we like to
show that diagrams of the type
\begin{equation}\label{classicaldiagram}
-\ii\Pi^{-+}_{\mathrm{cl}}=\unitlength6mm
\begin{picture}(14,1)\put(0,.2){\classdiagram}\end{picture}
\end{equation}
with $n$ $\{-+\}$ scattering interactions (intermediate particle-hole
$\{-+\}$ neutron loops) correspond to the $n$-th term in the Langevin
result (\ref{Mtau}). To demonstrate this we look into the time
structure of such a diagram, and assign times $0$ and $\tau$ to the
external $-$ and $+$ vertices, while the $-$ and $+$ interactions are
taken at $t^-_1$ to $t^-_n$ and $t^+_1$ to $t^+_n$, respectively. In
the classical limit $G^{--}$ is retarded, while $G^{++}$ is advanced
(see eqs. (\ref{RAgrin}), (\ref{tp}) and (\ref{tp2})), such that both
time sequences have the same time ordering: $0<t^-_1<\dots<
t^-_n<\tau$ and $0<t^+_1<\dots< t^+_n<\tau$ (all inequalities reverse,
if all line senses are reversed). Thus the $\tau$-dependence of the
modulus of these diagrams gives $e^{-|\Gamma\tau|}$. A second
simplification emerges from the fact that the $+-$ loop interaction
insertions mediate classical momentum transfers $|{\vec p}_n-{\vec
p}_{n+1}|$ which are large compared to $\hbar\Gamma$. Therefore the
time structure of these loops becomes very short on the scale
$1/\Gamma$ and therefore merge $\delta$-functions: $\delta
(t^-_n-t^+_n)$. With $t_n=t^-_n=t^+_n$ diagram
(\ref{classicaldiagram}) then no longer depends on the intermediate
times $t_n$ apart from the ordering condition, and therefore results
in a factor $|\Gamma\tau|^n/n!$ With $\hbar{\vec q}=0$ also the
corresponding momenta are pair-wise identical, and the remaining
momentum integrations just serve to define the correlation between
${\vec v}_m$ and ${\vec v}_{m+n}$ after $n$ scattering. Thus
\begin{eqnarray}\label{diagLangevin}
&&\unitlength6mm\begin{picture}(6.5,.7)\put(0,.2){\nloop}\end{picture}
  \\[3mm] &&\hspace*{2cm}= \left\{\begin{array}{ll}
  -\ii\Pi^{-+}_n (\tau ,{\vec q}=0)&\simeq 4\pi e^2 \rho_0\left<v^i_m
  v^k_{m+n}\right>_m \frac{|\Gamma\tau|^n}{n!} e^{-|\Gamma\tau|}\\
  -\ii\Pi^{-+}_n (\omega ,{\vec q}=0) &\simeq 4\pi e^2 \rho_0 \left<v^i_m
  v^k_{m+n}\right>_m 2\Gamma^n \Re\frac{1}{(\Gamma-\ii\omega)^{n+1}}
  \end{array}\right.,\nonumber
\end{eqnarray}
where the resulting proportionality to the proton density $\rho_0$
results from $-+$ and $+-$ Green's functions next to the external
vertices.  Here of course we have silently assumed a consistency
condition to be fulfilled: namely that the interaction loops which one
takes into account also consistently define the damping width $\Gamma$
of the source particles! In particular the extra powers in occupations
coming from the $+-$ loops are contained in $\Gamma$, and therefore do
no longer explicitly appear in the final result. This proves that in
the classical limit these diagrams reproduce the terms of the
classical Langevin series (\ref{Mtau}).

\subsection{Hierarchy of the QC expansion}
The Langevin diagrams (\ref{diagLangevin}) have the following
properties:
\newcounter{lst}
\begin {list}{\alph{lst})}{\usecounter{lst}}
\item {\em external vertices}: the external photon couples to the same
fermion line;
\item {\em topology}: the diagrams are planar, i.e. no crossing of
lines occur.
\item $-+$ {\em sign topology}: if one cuts them at all $-+$ lines
they decompose precisely into two pieces;
\item {\em value}: apart from the velocity correlations they all give
the same contribution to the soft photon point $(\omega,{\vec q})=0$.;
\end{list}
Using the equal time properties of classical $-+$-loop interaction
insertions one finds that
\begin{equation}\unitlength6mm\label{insert+-}
    \begin{picture}(2.8,1)\put(-0.3,.15){\put(.6,.75){\vector(1,0){1.8}}
    \put(2.4,-.75){\vector(-1,0){1.8}}
    \put(1,0){\interaction}\put(2.2,.95){\ssp}
    \put(2.2,-.95){\ssm}}\end{picture}
    =\int_0^\infty \Gamma e^{-\Gamma\tau}\di \tau=1,
    \quad{\rm while}\quad
    \begin{picture}(2.8,1)\put(-0.3,.15){\put(.6,.75){\vector(1,0){1.8}}
    \put(2.4,-.75){\vector(-1,0){1.8}}
    \put(1,0){\interaction}\put(2.2,.95){\ssm}
    \put(2.2,-.95){\ssp}}\end{picture}
    =\int_0^\infty n \Gamma e^{-\Gamma\tau}\di \tau=n
\end{equation}
at ${\vec q}=0$, where $n$ is the proton occupation.  The left case
iteratively enters the Langevin diagrams and one compiles
factors of unity for each time folding, since the loop insertions of
order $\Gamma$ are compensated by the time integration over the
Green's function product $G^{--}G^{++}$. This proves d).

What remains to be shown is that for the classical problems discussed
in sect. \ref{sect-class} all other key diagrams are disfavored by extra
occupation factors $n$ or $\Gamma/\bar{\epsilon}$ which both are small
compared to unity.

To a): Since in the classical problems of sect. \ref{sect-class} only
a single charged particle (say a proton) in neutral matter is
considered, all diagrams with more than one proton line do not occur
in this case. To b): Non-planar diagrams, where interactions cross,
violate classical time-ordering. Either the two interaction times are
interlocked on a time scale $1/\bar{\epsilon}$ and therefore lead to a
penalty factor $\Gamma/\bar{\epsilon}$ or restoring the time-ordering
one obtains a Z-shape fermion line, which gives an extra occupation
factor $n$.  To c): Diagrams that can be cut into more than two pieces
can be obtained on various ways: i) by any $-+$-loop insertion which
switches signs on the outer proton lines, thus using the right
insertion in (\ref{insert+-}); here one obtains an extra hindrance
factor $n$.  ii) extra insertions of $--$ or $++$ blocks linking the
outer proton lines; since there are no direct proton-proton
interactions these effective interactions are mediated by the
surrounding neutron matter thus containing $--$ neutron loops, which
also leads to an additional factor $n$, besides another $n$ factor due
to the $-+$ proton Green's function.

\section{General Quantum Consideration for Hot and Dense
Matter}

The considerations above show the following. A proper treatment of an
entirely classical problem, namely the coupling of a classical source
to a wave (electromagnetic field), on the level of quantum many-body
theory requires technics, that even nowadays are still non-standard,
i.e.  beyond perturbation theory or QPA. While the classical problem
can be solved quite conveniently and simple with no problems on the
infra red side, the corresponding quantum description requires an
appropriate account of the finite damping width $\Gamma$ of the source
particles. The most natural approach in our mind is the real-time
Green's function technic, which however requires partial resummation,
such that the finite width is included already on the one-body Green's
function level.

In this section we analyze the production rate from hot and dense
matter in the quantum case in terms of non--equilibrium Green's
functions.  In order to provide some analytical results which easily
can be discussed in different limiting cases, we employ the following
approximation for the full retarded Green's function. We assume $G^R$
to be given by a simple pole approximation with constant residue
\begin{equation}
G_R^F=\frac{1}{\epsilon+\mu_F-\epsilon({\vec p})+\ii\Gamma/2}
\end{equation}
where $\epsilon({\vec p})=\vec{p}^2/(2m^{\ast}_{F})$, $m^{\ast}_{F}$
being an effective fermion mass. The width $\Gamma$ is assumed to be
independent of $\epsilon$ and ${\vec p}$. Explicit results will be
given for the one - and three loop case (the first two diagrams in
(\ref{keydiagrams})).

\subsection{Qualitative expectations}
Compared to the classical results we expect the following changes:
\newcounter{lst1}
\begin{list}{\alph{lst1})}{\usecounter{lst1}}
\item one looses the classical hierarchy of diagrams, such that many
more diagrams contribute in the quantum case; for specific couplings,
however, some diagrams are disfavored or drop due to selection or
suppression rules, e.g. non-planar diagrams in SU(n) coupling;
\item the radiated quantum carries finite momentum and energy (which
vanish in the classical limit), such that additional recoil
corrections and phase-space factors $\sim e^{-\omega/T}$ appear; the
latter is important, since it cures the classical ultra violet
catastrophe, where the intensity spectrum is white, leading to a
divergence of the radiated energy in the classical case;
\item the occupations are no longer of Boltzmann type with $n\ll
1$, so that Pauli suppression and Bose-Einstein enhancement effects
are significant;
\item the duration time of binary collisions $\sim 1/\epsilon_F$
mediated by interaction loop insertions (\ref{open}) in the
correlation diagrams, are no longer negligible compare to $1/\Gamma$
as in the classical case, so that non-markovian memory effects become
important \cite{CGreiner,MR}.
\end{list}
Points (b) and (c) can be clarified using the exact relation between
thermal fermion and boson occupations
\begin{eqnarray}
\label{nrel}
  n^F(\epsilon+\omega/2)(1-n^F(\epsilon-\omega/2))
  &=&(n^F(\epsilon-\omega/2)-n^F(\epsilon+\omega/2))n^B(\omega)
\end{eqnarray}
which simplifies in the following limits to
\begin{eqnarray}
n^F(\epsilon+\omega/2)(1-n^F(\epsilon-\omega/2))
  &\approx&-\frac{\di}{\di\epsilon}n^F(\epsilon)n^B(\omega)\omega \for
  \omega\ll T\\
  &\approx&n^F(\epsilon)n^B(\omega)\omega /T\for
  \omega\ll T , n^F\ll 1\nonumber.
\end{eqnarray}
At low temperatures only states close to the Fermi-surface
contribute. The last approximate relation suggests that relative to
the classical results of sect. 3 an additional phase-space suppression
factor $n^B(\omega)\omega/T$ appears in the quantum case, which
accounts for the finite energy $\omega$ carried away by the quantum.

\subsection{Contribution of One--Loop Diagram with Full
Fermion Propagators}

We first consider the one--loop diagram of (\ref{keydiagrams}) with
the full fermion propagators
\begin{equation}\label{one-loop}
\unitlength6mm\begin{picture}(3.5,1)
   \put(0,0.15){\oneloopvertex}\end{picture}
   =-\ii\Pi^{-+}_0=-\ii V^{\mu} V^{\nu} {\cal A}_0^{-+}
\end{equation}
where ${\cal A}_0^{-+}$ denotes the bare loop without vertices.  For
simplicity we will neglect vertex corrections.  The later can be
trivially included in Landau--Migdal approximation (e.g.  see
eq. (\ref{onelooptau}), Appendix B and refs. \cite{MSTV,Voskr}).

With the help of relation (\ref{nrel}) and the equilibrium
form of the Green's functions (\ref{Geq}) the bare loop
reads
\begin{equation}
\hspace*{-5mm}\label{A-+}
-\ii{\cal A}_0^{-+}=
n^B_{\omega}\int \frac{\di\epsilon
\di^3p}{(2\pi)^4}\frac{\Gamma}{(\epsilon+\mu_F-\epsilon_p)^2+(\Gamma/2)^2}
\frac{\Gamma \,\, (n^F_{\epsilon}-n^F_{\epsilon+\omega})}
{(\epsilon+\mu_F-\epsilon_{\vec{p}+\vec{k}}+\omega)^2+ (\Gamma/2)^2}.
\end{equation}

This expression can be evaluated in closed form in different limits.
We first analyze the QPA ( i.e. $\Gamma \ll\omega, kv_F,T$ in case
$T\ll\epsilon_F$ or $\Gamma\ll\omega , kv_T\sim k\sqrt{T/m^{\ast}_F}$,
for $T\gg\epsilon_F$), which can be found in the literature. In this
limit one recognizes two energy $\delta$--functions in eq. (\ref{A-+})
(see approximation formula (\ref{imquasiG}) of Appendix A), which
together with momentum conservation and exact relativistic kinematic
can only be fulfilled for space-like $(\omega ,{\vec k})$. Following
ref. \cite{MSTV} one obtains for the QPA loop in various limits
\begin{eqnarray}
\label{ImAT}\hspace*{-10mm}
&-\ii&\left.{\cal A}_{0}^{-+}(\omega, \vec{k})\right|_{\rm{QPA}}=
\left\{\begin{array}{ll}0&\for \omega\ge|{\vec k}|\\
\frac{\mbox{$(m^{\ast}_F)^2  T$}}{\mbox{$2\pi k$}}n^B_{\omega}
\ln\frac{\mbox{$\exp(\kappa)+1$}}{\mbox{$\exp(\kappa)
+\exp(-\omega/T)$}}&\for \omega\ll|{\vec k}|,\end{array}\right.\cr
&\rightarrow&
\left\{\begin{array}{lll}
\frac{\mbox{$(m^{\ast}_F)^2  T$}}{\mbox{$2\pi k$}}n^B_{\omega}
&\for\kappa \ll -1,\quad &\omega\ll|{\vec k}|\\
\frac{\mbox{$(m^{\ast}_F)^2  T$}}{\mbox{$2\pi k$}}
\mbox{$\exp(-\kappa)\exp(-\omega/T)$}
&\for \kappa \gg 1,\quad&\omega\ll|{\vec k}|,\end{array}\right.
\\[5mm]\nonumber&&\where
   \kappa=(m^{\ast}_Fv^2/2-\mu_F)/T,
   \with v= (\omega-k^2/(2m^{\ast}_F))/|{\vec k}|.
\end{eqnarray}
Here non-relativistic kinematics\footnote {As a defect of the
non-relativistic approximation the result does not exactly vanish for
$\omega\ge|{\vec k}|$ but rather leads to terms of the order
$\exp(-m^{\ast}_F/T)$ or less.} has been used, where $v$ is the recoil
corrected fermion velocity that essentially contributes to the loop
and the condition $\omega\ll|{\vec k}|$ assures $|v|\ll 1$. The
simplified expression for $\kappa \ll -1$, realized for
$T\ll\epsilon_F\approx \mu_F$ and $\omega < kv_F$, is quite frequently
used for space-like interaction loops in low temperature Fermi systems
as in Landau's Fermi liquid theory. However, it shows a singular
behavior $\propto 1/(k\omega )$ in the small $\omega\ll k=|{\vec
k}|\ll T$ limit, which is a generic defect of the QPA. The Boltzmann
limit $\mu_F\ll -T$ leads to the $\kappa\gg 1$ case, which apart from
recoil correction and the extra quantum phase-space factor
$e^{-\hbar\omega/T}$ coincides with that for classical diffusion
result (\ref{Pidifflq}) in sect. 3. In summary, for the one loop term
the QPA leads to meaningful results only for large space-like
$\omega,{\vec q}$ (hard thermal loops).

In the general case only the integral in eq. (\ref{A-+}) over the angle
$\hat{p}\hat{k}$ can be performed
\begin{eqnarray}
\label{ImAarc}
-\ii{\cal A}_0^{-+}&=&n^B_{\omega}\int_{-\infty}^{\infty}\di\epsilon
 \int_{0}^{\infty}\frac{
p\di pm^{\ast}_F}{(2\pi)^3k}\frac{\Gamma}{(\epsilon+\mu_F-\epsilon_p)^2
+(\Gamma/2)^2}
(n^F_{\epsilon}-n^F_{\epsilon+\omega})
\cr
&\times& \left[\arctan\frac{\epsilon
+\mu_F+\omega-\epsilon_p-\epsilon_k+pk/m^{\ast}_F}{(\Gamma/2)}\right.\\
&-&\left.\arctan\frac{\epsilon
+\mu_F+\omega-\epsilon_p-\epsilon_k-pk/m^{\ast}_F}{(\Gamma/2)}\right]
\nonumber
\end{eqnarray}
in closed form. To proceed further we consider the case of small
spatial momenta $k=|{\vec k}|$ such that $pk\ll m^{\ast}_F\Gamma$. For
the remaining two-dimensional integral
\begin{eqnarray}
 \label{Imsimp}
  -\ii{\cal A}_0^{-+}&=& \frac{(2m_F^{\ast})^{3/2}}{(2\pi)^3}n^B_{\omega}
  \int_{-\infty}^{\infty}
  \di\epsilon(n^F_{\epsilon}-n^F_{\epsilon+\omega})
  \int_{0}^{\infty} \di\epsilon_{p}(\epsilon_{p})^{1/2}
  \frac{\Gamma}{(\epsilon+\mu_F-\epsilon_p)^2+(\Gamma/2)^2}
\cr&\times&  \frac{\Gamma}{(\epsilon+\mu_F-\epsilon_p-\epsilon_k+\omega)^2+
 (\Gamma/2)^2} \for pk\ll m^{\ast}_F\Gamma
\end{eqnarray}
one realizes that the $\epsilon_p$-integration over the product of
Lorentz functions gives
\begin{equation}
\frac{4\pi\Gamma\sqrt{\epsilon+\mu_F}}{(\omega-k^2/(2m^{\ast}_F))^2+\Gamma^2}
\for \epsilon+\mu_F\gg\Gamma
\end{equation}
while it is essentially zero for $\epsilon+\mu_F\ll -\Gamma$. For
correspondingly small values of $\Gamma$ one therefore obtains
\begin{equation}
 \label{KB}
 -\ii{\cal A}_0^{-+}\simeq n^B_{\omega}
 \frac{\Gamma (2m^{\ast}_F )^{3/2}}{2\pi^2
 [(\omega -k^2/2m^{\ast}_F)^2+\Gamma^2]}\int_0^\infty \epsilon_p^{1/2}
 d\epsilon_p (n^F_{\epsilon_p-\mu_F}-
 n^F_{\epsilon_p+\omega-\mu_F}),
  \end{equation}
where we have replaced the remaining $\epsilon$ variable by $\epsilon_p$.
The very same form emerges, if one formally replaces
\begin{equation} \label{ap3}
 n^F_\epsilon - n^F_{\epsilon+\omega}\rightarrow n^F_{\epsilon_p-\mu_F}-
n^F_{\epsilon_p+\omega-\mu_F}
\end{equation}
in (\ref{Imsimp}) and first integrates over $\epsilon$.  This approximation
(\ref{ap3}) corresponds to the Kadanoff--Baym ansatz (see Appendix A).

Equation (\ref{KB}) valid for $pk\ll
m^{\ast}_F\Gamma$ can be evaluated in two limits
\begin{equation}\label{omgam}
\hspace*{-1cm}-\ii{\cal A}_0^{-+}\longrightarrow
 \frac{\omega}{T}n^B_{\omega}
\frac{\Gamma}{(\omega-k^2/(2m_F^{\ast}))^2+\Gamma^2}
\left\{\begin{array}{ll}
     m_F^{\ast}p_FT/\pi^2&\for \Gamma,T\ll \epsilon_F\approx\mu_F\\
     \rho_F&\for T\gg\epsilon_F,\omega,\Gamma\end{array}\right.
\end{equation}
where essential contributions arise from momenta $\sim p_F$ for
$T\ll\epsilon_F$ or around $p_T\sim\sqrt{m^{\ast}_FT}$ for
$T>\epsilon_F$. Here $\rho_F$ is the density of the charged
fermions. Compared to the QPA which is zero for time-like momenta,
this result is finite and of order $1/\Gamma$ in the soft limit. It
agrees with the classical result (\ref{Mtaun0}) and the corresponding
QC limit (\ref{oneloopomega}) besides recoil and the quantum
phase-space corrections.

Starting from the QPA for the fermion Green's functions one usually
attempts to restore a dependence on the non-zero fermion width for the
boson self--energy $\Sigma_{B}^{R}$ by means of the analytical
continuation $\omega
\rightarrow \omega+\ii\Gamma$, where
\begin{equation}
\label{Aan}
 {\cal A}^{R}_0=\int
\frac{d^3p}{(2\pi)^3}\frac{n_p-n_{\vec{p}+\vec{k}}}{\epsilon_p-
\epsilon_{\vec{p}+\vec{k}}+\omega+\ii\Gamma}\,.
\end{equation}

In refs. \cite{V2,MSTV,HU} such a procedure has been used in order to
account for the finite $\Delta$--isobar width in the pion self energy.
We see that the value $\Im {\cal A}^{R}_0$ given by eq. (\ref{Aan}) and
that given by (\ref{KB}) coincide only if $n_{\vec{p}+\vec{k}}\simeq
n_{\epsilon_p+\omega-\mu_F}$, i.e. applying Kadanoff--Baym ansatz.

The thus discussed one--loop diagram can also be used for the
intermediate $t$-channel interaction loops ${\cal G}^{-+}$ occuring in
higher order diagrams. There the typical values of parameters are
\begin{equation}
\label{smallom}
 \omega < \epsilon_F\,,\,\,k\geq p_F\,,\,\,p\geq p_F\;.
\end{equation}
Therefore at least for $\Gamma \ll\epsilon_F$ one has large space-like
momenta with $pk\gg\Gamma m^{\ast}_F$. Then from eq. (\ref{ImAarc}) one
obtains
\begin{equation}
\label{nularge}
 -\ii{\cal A}^{-+}_0=-\int_{-\infty}^{\infty}
 \frac{(m^{\ast}_F)^2}{4\pi^2k}(n^F_{\epsilon}-
n^F_{\epsilon+\omega})d\epsilon\,
[\frac{\pi}{2}+\mbox{arctan}(\frac{\epsilon+\mu_F}{\Gamma})]\,,
\end{equation}
which merges the QPA expressions in (\ref{ImAT}) for
large space-like momenta in the limit $\Gamma\ll \epsilon_F$ both at
low temperatures $T\ll\epsilon_F\approx\mu_F$ ($\kappa\ll-1$) and in the
Boltzmann limit ($\kappa\gg1$).

Thus, for the intermediate $t$-channel interaction loops ${\cal
G}^{-+}$ one can safely use the QPA ($\Gamma\rightarrow 0$), which
even accounts for higher order correction to this loop, as one
sees from the corresponding limit of the diffusion result
(\ref{Pidifflq}). Only for soft loops (e.g. as in the $s$-channel) the
QPA is ill defined.

Comparing the one-loop result at non-zero $\Gamma$ (\ref{omgam}) with
the first non-zero diagram in the QPA ($\Gamma=0$ in the fermion
Green's functions)
\begin{equation}\label{corr0}
\unitlength8mm\begin{picture}(3.5,1)
   \put(0,0.15){\oneloopvertex}\end{picture}
   = C_0(\omega)
   \left\{\begin{picture}(3.5,1)\put(0,.15){\selfinsert}\end{picture}
   \right\}_{\mbox{QPA}}
\end{equation}
at small momentum ${\vec k}$
one determines a correction factor
\begin{equation}
\label{C}
  C_0(\omega)=\frac{\omega^2}{\omega^2+\Gamma^2}\,,
\end{equation}
which cures the defect of the QPA for soft $\omega$. This factor
complies with the replacement $\omega\rightarrow \omega+\ii\Gamma$. A
similar factor has been observed in the diffusion result, where
however the macroscopic relaxation rate $\Gamma_x$ enters, due to the
resummation of all rescattering processes. Other factors between
eq. (\ref{omgam}) and in the corresponding QPA Feynman diagram (c.f.
ref. \cite{V3}) become identical, if one explicitly calculates the
width $\Gamma$ to that order, see also \cite{V2}.

\subsection{Higher Order Diagrams with Full Fermion Propagators}

Along similar routes (details are given in Appendix C) the correction
factors for the higher order diagrams can be derived. Here we just
quote the results for the next lowest order diagrams
\begin{eqnarray}
  \unitlength6mm\begin{picture}(3.5,1.)
  \put(0,0.15){\oneloopvertex}\put(1.625,0.15){\interaction}
  \end{picture}&=&C_1(\omega)\left\{\;
  \unitlength6mm\begin{picture}(3.5,1)
  \put(0,0.15){\oneloopvertex}\put(1.625,0.15){\interaction}
  \end{picture}\right\}_{\mbox{QPA}}\label{oneint}\label{corr1}\\
  \cr \hspace*{1cm}&&\cr
  \unitlength6mm
  \begin{picture}(3.8,1.)
  \put(0,0.15){\doubleloop}\end{picture}&=&C_0(\omega)
  \left\{\unitlength6mm \begin{picture}(3.8,1.)
  \put(0,0.15){\doubleloop}\end{picture}\right\}_{\mbox{QPA}}\label{corr2}
\end{eqnarray}
with $C_0(\omega)$ from (\ref{C}) and
\begin{equation}
\label{C1}
  C_1(\omega)=\omega^2\frac{\omega^2-\Gamma^2}{(\omega^2+\Gamma^2)^2}\,.
\end{equation}
The total radiation rate is obtained from all
diagrams in (\ref{keydiagrams}).

The set of QPA diagrams in (\ref{corr0}), (\ref{corr1}) and
(\ref{corr2}) are just those that determine the IQF scattering rate
(\ref{MIQF}) including the exchange diagram for a source of fermions
(\ref{corr2}). The latter drops in the classical limit, where the two
other ones yield the $n=0$ and $n=1$ terms of the classical Langevin
result (\ref{Mtau}). Thereby the damping correction factors $C_0$ and
$C_1$ in the quantum case are the same as classically derived.  For
$\Gamma \ll \omega$ they tend to unity and the production rate
coincides with the QPA results as obtained in ref. \cite{V3}, while
there is a substantial suppression at small frequencies
$\omega\ll\Gamma$.

In the corresponding QC limit all the diagrams of type
(\ref{classicaldiagram}) with an arbitrary number of $-+$
$NN$--interaction insertions can be summed up leading to the diffusion
result (\ref{Mclres}) in sect. 3.1. For small momenta ${\vec q}$ this
leads to a suppression factor of the form
$C={\omega^2}/{(\omega^2+\Gamma_x^2)}$.

There is hope that even in the quantum case some higher order
diagrams can also be resummed and that qualitatively a similar
suppression factor emerges like for the diffusion result.

\section {Discussion and Perspectives}

We investigated the production of particles from the collision
dynamics of dense matter at the example of photon production. Thereby
the source of charged particles was described in two ways, a) as a
classical system governed by classical transport equations and b) as a
quantum system in terms of a real-time non-equilibrium field theory
formulation. The central quantity is the current-current correlation
function which relates to the imaginary part of the proper self energy
of the produced particle. Under quite general assumptions this
correlation function governs the local production and absorption rates
in the matter. Since the here discussed features are of kinematical
origin, relating space-time scales to the corresponding momentum and
energy scales, all conclusions drawn in this study are general and
therefore also apply to the in-medium production and absorption rates
of any kind of particle.

The problem could be quite naturally formulated and solved in the
classical descriptions by means of a macroscopic transport and a
microscopic Langevin process. These studies showed that spectrum of
produced particles is essentially governed by one macroscopic scale,
the relaxation rate $\Gamma_x$ of the source. For frequencies $\omega$
of the produced quantum which are large compared to $\Gamma_x$ the
spectrum can be described by the incoherent quasi-free scattering
approximation (IQF) used in most of the transport models. Higher order
corrections help to improve the result. Once, however,
$\omega<\Gamma_x$ this quantum is ''soft''. It can no longer resolve
the individual collisions in time and therefore the IQF picture fails
and produces a false infra-red divergence in the rate. Rather, the
correct rate is regular and differs from the IQF result by a
suppression factor
\begin{equation}\label{suppression}
C_0(\omega)=\frac{\omega^2}{\omega^2+\Gamma_x^2}\; .
\end{equation}
This soft part of the spectrum is genuine non-perturbative. The
essential features are summarized in figs. 1 and 2 of sect. 3.  For
relativistic sources a second scale comes in, once the wave number
$|{\vec q}|$ of the produced particle exceeds the value of
$\Gamma_x/\sqrt{\left<v^2\right>}$ due to the increased spatial
resolution as shown by the closed form results of the diffusion model.

On the quantum level new scales come in since now $\omega$ and ${\vec
q}$ also correspond to the energy and momentum of the photon which
have to be compared with the characteristic energy and momentum scales
of the source as given by temperature $T$ and chemical potential $\mu$
or the Fermi energy $\epsilon_F$. Also the occupations can become
degenerate and Pauli suppression or Bose enhancement effects are
important. For the general formulation one has to leave theoretical
schemes that are based on the concept of asymptotic states like
perturbation theory or quasi-particle approximation. The proper frame
is the real-time non-equilibrium field theory, where the proper
self-energy can be formulated in terms of closed correlation diagrams
with general propagators. Thereby the resummation of Dyson's equation
to full propagators which also include the imaginary parts of the self
energy and therefore account for the damping of the source particles
is the essential step to cure the infra-red problem. Only this way one
comes to a convergent scheme. After this resummation the corresponding
set of diagrams is then reduced to diagrams with skeleton
topology. Using the Keldysh $-+$ notations we have addressed a
particular role to all $-+$ and $+-$ lines as 8-dimensional Wigner
densities of occupied and available ''states''. This motivated further
resummations which define in-medium interactions and vertex
corrections. The resulting set of diagrams can then be discussed in
detail in these physical terms. A decomposition of these correlation
diagrams in terms of the interference of two ''amplitude'' diagrams is
suggested, where Wigner densities enter as in- and out-states. This
permits a transparent physical interpretation of the correlations
diagrams, which may be used to formulate multi-particle collision
processes in matter. They can serve as input for a generalized transport
description, which ultimately includes the off-shell propagation of
particles and therefore unifies resonances which have a width already
in vacuum with all other particles in the dense matter, which acquire a
damping width due to collisions.

Once one has the full propagators and the in-medium interactions it
is in principle straight forward calculate the diagrams.  However both,
the computational effort to calculate a single diagram and the number
of diagrams, are increasing dramatically with the loop order, such
that in practice only lowest order loop diagrams can be considered in
the full quantum case. In certain limits some diagrams drop out. In
particular we could show, that in the classical limit of the quantum
description only a special set of diagrams survive, which could be
associated with the multiple collision terms of the classical random
Langevin process. Comparing the lowest order loop diagrams in various
limits to the corresponding QPA diagrams one realizes that also here
correction factors similar to (\ref{suppression}) appear. Now the
characteristic scale is the damping width $\Gamma$ of the source
particles. Accounting for higher order diagrams one concludes that
also in the quantum case the relaxation rate $\Gamma_x$ is the
relevant scale which decides between soft and hard photons.
Thus for applications one has to compare the typical energies of the
produced particles with the typical relaxation rates of the source
system.

Our considerations are of particular importance for the theoretical
description of nucleus-nucleus collisions at intermediate to
relativistic energies. With temperatures $T$ in the range of 30 to 100
MeV for dense nuclear matter, up to 200 MeV for hadronic matter and
beyond 150 MeV for the quark gluon plasma or parton phase most of the
kinetic models that are used infer collision rates $\Gamma$ for the
constituents, which during the high density phase can reach the
system's temperature, $\Gamma\lsim T$.  Such estimates make the use of
on-shell concepts already rather questionable. The particles
uncertainty in energy is comparable with the mean kinetic energy! In
particular the bulk production and absorption rates of all particles
with masses less than $T$, if calculated in standard IQF
approximation, are seriously subjected to the here discussed
effect. Therefore the corresponding quenching factors
(\ref{suppression}) should sensitively affect the production rates of
quark pairs and gluons during the plasma phase, of low energy pions
during during hadronization and real and virtual photons with
correspondingly low energies. Since our discussion was restricted to
the production in dense matter, for the particular case of photon
production in nuclear collisions one has to consider in addition the
radiation caused by the incoming charged ions and outgoing charged
fragments. Due to Low's theorem \cite{Low} the latter give rise to an
infra-red divergent $\sim 1/\omega$ component which interferes with
the one discussed here.

In astrophysics neutrinos produced from neutron stars or during super
nova collapse have an absorption mean free path which is long compared
to the size of the radiating system (for review see
\cite{ST,NS,MSTV}). Thus the production rates cannot be estimated by
black-body radiation. Rather the microscopic rates are relevant. The
mean kinetic energies per neutrino or $\nu{\bar \nu}$-pair are about
$\sim 3T$ or $\sim 6T$, respectively \cite{FM,SS}.  With mean
collision rates of the order of $\Gamma\sim \pi^2T^2/\epsilon_F\ll T$
\cite{MorelNozieres}, c.f. eq. (\ref{sigret}), the production rates
can safely be estimated in IQF-approximation for relatively cold
neutron stars $T\leq 1.5$ to 2 MeV. Already around $T\sim 5$ MeV the
quenching factor (\ref{suppression}) is significant ($0.3$) and it may
become even smaller during super nova collapse. Then the temperatures
can raise to $T\sim 10-30$ MeV such that $\Gamma\sim T$ and the here
discussed suppression effects are relevant for the corresponding
neutrino emissivity.

\appendix{\bf Appendix}
\section{Kadanoff--Baym Ansatz and QPA}

In the general non--equilibrium case one has no simple relations between
Green's functions as in equilibrium. In order to proceed nevertheless
one often uses the so called Kadanoff--Baym ansatz \cite{KB}.  For
Fermions it reads
\begin{equation}\label{KBa}
\label{anz}
G^{+ -}_F =2\ii (1-n^F_{\epsilon_{\vec p}-\mu_F})\Im G^{R}_F ,\quad
G^{- +}_F =-2\ii n^F_{\epsilon_{\vec p}-\mu_F}\Im G^{R}_F ,
\end{equation}
where $n^F_{\epsilon_p-\mu_F}$ are the fermion occupations which now
depend on $\vec{p}$ through the on-shell dispersion relation
(\ref{ep}) rather than on $\epsilon$. One should note that the
Kadanoff--Baym ansatz does not directly follow from the properties of
the $G^{- +}$ and $G^{+ -}$ functions, rather it has been introduced in
order to recover the Boltzmann limit. The correctness of this ansatz
has only been proven in the QPA, see \cite{SL}. Eqs. (\ref{KBa})
complies with the definition of the particle densities
(\ref{part-dens}), as can be seen by the sum rules (\ref{SRA}).

Dealing with dressed particles we consider only diagrams with thick
fermion lines determined from the corresponding Dyson
equations. Approximation ({\ref{KBa}) is however based on the
assumption that $\Im \Sigma^{R}_F$ is much smaller than all other
energies scales entering the problem.

In particular in the limit $\Im \Sigma^{R}_F \rightarrow 0$ in the
fermion Green's functions one comes to the QPA, where the imaginary
part of retarded Green's function becomes a delta function, e.g. for
non--relativistic fermions
\begin{equation}
\label{imquasiG}
\Im G_F^{R}\simeq -\frac {\pi}{Z_{\vec p}} \delta
[\epsilon + \mu_F - \epsilon_{\vec p}^0 -\Re \Sigma^{R}_F
(\epsilon +\mu_F ,\vec{p})]\,\,,\quad Z_{\vec{p}}=1/(1 - {\partial
\mbox{Re}\Sigma^R}/{\partial \epsilon})
\end{equation}
where the dispersion relation between $\epsilon$ and $\vec{p}$
is implicitly given by
\begin{equation}
\label{eps}
 \epsilon_{\vec p}\simeq \epsilon^0_{\vec p}+\Re \Sigma^{R}_F
 (\epsilon_{\vec p}, \vec{p}).
\end{equation}
Here the $Z_{\vec p}$ factor corrects for retardation effects,
c.f. ref. \cite{VBRS}, such that the sum rule (\ref{SRA}) remains
fulfilled. For simplicity we employ a quadratic $p$--dependence for
$\Re \Sigma^R$ in terms of an effective fermion mass $\epsilon_{\vec
p}\simeq\vec{p}^2/2m^{\ast}_{F}$ in the applications.

{}From the dispersion equations one easily finds the corresponding
relaxation times. E.g., for fermions supposing that $\Im \Sigma^{R}$
is small and introducing $\tau_{\mbox{col}}=1/2\delta \epsilon $,
where we use that $\psi_F \sim \mbox{exp}(-\ii(\epsilon_{\vec
p}-\ii\delta \epsilon)t)$, one finds
\begin{equation}
\label{taun}
\tau_{\mathrm{col}} =\left\vert
\left( 1-\frac{\partial \Re \Sigma^R_F }{\partial
\epsilon}\right)/\left(2\Im \Sigma^{R}_F \right)
\right\vert_{\epsilon=\epsilon_{\vec p}}\,.
\end{equation}
This value would tend to infinity for $\Im \Sigma^{R}_F \rightarrow
0$.  In reality the fermion width determined by the value $\Im
\Sigma^{R}_F$ is rather large even at sufficiently small
temperatures. E.g. for nucleons, applying the QPA for the intermediate
nucleon lines, it can be estimated as follows \cite{MorelNozieres}
\begin{equation} \label{sigret}
\Im \Sigma^{R}_F \simeq -\vert \tilde{M}_F
\vert^2[(1-\epsilon_{\vec p}/\epsilon_F)^2+T^2\pi^2/\epsilon^2_F],
\quad T\ll \epsilon_F
\end{equation}
with typically $\vert \tilde{M}_F\vert^2\sim \epsilon_F$ at
normal nuclear density \cite{V2}. This estimate shows that
$\vert\Im \Sigma^{R}_F \vert$ comes into the order of $\epsilon_F$
already at sufficiently small temperatures
$T\simeq\frac{1}{3}\epsilon_F$ and still increases for higher
$T$. This defers the application of the QPA for a wide range of
temperatures.

\section{Renormalization of the Two--Fermion Interaction}\label{AppB}

As an example we consider a theory where non-relativistic fermions
interact via two-body potentials. A priori this theory has no bosons
and the two-body interactions always connect two vertices of same
sign, defining $iV^{--}$ and $iV^{++}=\left(iV^{--}\right)^\dagger$,
while $V^{-+}=V^{+-}=0$.  Even if resumed to an effective four point
interaction ${\cal G}_0^{--}$ according to eq. (\ref{4point}) much like
Bruckner $G$--matrix, one has an effective interaction
that connects only like sign vertices.

For the following we approximate ${\cal G}_0$ by a two--point function
(as for instance in Fermi-liquid theory, where the residual
interaction is supposed to be local and extracted from comparison with
experimental data \cite{M,MSTV}), while ${\cal G}_0^{-+}={\cal
G}_0^{+-}=0$.  We also suppose that ${\cal G}_0^{++}$ and ${\cal
G}_0^{--}$ interactions are particle-hole irreducible in the
$t$-channel (vertical in (\ref{4point}), c.f. \cite{M,MSTV}).

Starting from ${\cal G}_0^{--}$ and ${\cal G}_0^{++}$
one can completely bosonize the interaction in the standard way by
resumming all intermediate particle-hole loop insertions
\begin{equation}
 {\cal A}_{12}^{ij}=G_{12}^{ik}G_{21}^{kj},\quad i, j\in \{- +\}
\end{equation}
through the Dyson equation in two by two matrix form
\begin{eqnarray}\label{effboson}
  &&\hspace*{1.8cm}{\boldsymbol {\cal G}}={\boldsymbol {\cal G}_0}
   +{\boldsymbol {\cal G}_0}\odot{\boldsymbol {\cal A}}
  \odot {\boldsymbol {\cal G}}.
\end{eqnarray}
In space-time homogeneous cases eq. (\ref{effboson}) can be solved
algebraically. First one defines a residual interactions through
repeated ${\cal A}^{--}$ and ${\cal A}^{++}$ insertions as
\begin{equation}
{\cal G}^{--}_{res}=\frac{{\cal G}^{--}_0}{1-{\cal G}^{--}_{0}
{\cal A}^{--}},\,\,
{\cal G}^{++}_{res}=\frac{{\cal G}^{++}_0}{1-{\cal G}^{++}_{0}
{\cal A}^{++}}=-\frac{{\cal G}^{--}_0}{1+{\cal G}^{--}_{0}
{\cal A}^{++}}
\end{equation}
Using ${\cal G}_0^{-+}={\cal G}_0^{+-}=0$  straight forward
algebra yields
\begin{eqnarray}\label{renormG}
{\cal G}^{+-}=&Z{\cal G}^{++}_{res}{\cal A}^{+-}{\cal G}^{--}_{res},\quad&
{\cal G}^{-+}=Z{\cal G}^{--}_{res}{\cal A}^{-+}{\cal G}^{++}_{res},
\cr
{\cal G}^{++}=&Z{\cal G}^{++}_{res},\quad&
{\cal G}^{--}=Z{\cal G}^{--}_{res},
\end{eqnarray}
for the components of the full interaction
${\boldsymbol {\cal G}}$, where
\begin{equation}\label{Zres}
Z=\left(1-{\cal G}^{++}_{res}{\cal A}^{+-}{\cal G}^{--}_{res}
{\cal A}^{-+}\right)^{-1}
\end{equation}
is a renormalization factor.

The explicit dependence of ${\cal G}^{-+}$ and ${\cal G}^{+-}$ on $Z$
can be moved to a renormalization of the loop
\begin{eqnarray}\label{renorm}
{\cal G}^{-+}={\cal G}^{--}_{res}{\cal A}_{ren}^{-+}{\cal
G}^{++}_{res},
\quad\mbox{where}\quad {\cal A}^{-+}_{ren}=Z{\cal A}^{-+}.
\end{eqnarray}

This full interaction ${\boldsymbol {\cal G}}$ has bosonic features,
just describing effective bosons, such as phonons, plasmons, sigma
mesons, etc.  Also the inclusion of real bosons, like pions in nuclear
matter, is possible, giving rise to a picture, where pions couple to
pionic particle-hole excitations, see \cite{Voskr,MSTV}.  These
effective bosons can be taken on the same footing as all other
effective quanta, the fermions or other bosons, see
\cite{V4,V1}. Thus, effective bosons also acquire a spectral function
with width and the non-diagonal components of ${\cal G}^{- +}$ and
${\cal G}^{+ -}$ are also Wigner functions.  Consequently one comes to
a theory of effective in-medium fermions interacting with effective
in-medium bosons.

Obviously the full (anti-)time-ordered ${\cal G}^{--}$ (${\cal
G}^{++}$) depend on the in-matter densities $G^{-+}$ and $G^{+-}$ also
via $Z$. However in an approximation where the value ${\cal G}^{++}_0
{\cal A}^{+-} {\cal G}^{--}_0 {\cal A}^{-+}$ is small, one has
$Z\simeq 1$ for renormalization factor (\ref{Zres}).  So, one can
simplify further and comes to a scheme like leading logarithmic
approximation in quantum field theory, namely, a perturbation series
over ${\cal G}^{- +}$ (or ${\cal G}^{+ -}$) neglecting corrections
$1+O({\cal G}^{+ -}{\cal A}^{-+})$ in each leading term. Such
corrections are proportional to $\rho^2$ (or $\Gamma^2$). Thus, one
approximately has
\begin{equation}\label{open}\unitlength6mm\label{4pointint}
    {\cal G}^{-+}=
    \begin{picture}(1.4,1)\put(-0.3,.15){\thicklines\put(1,0){\fullbox}
    \put(1,0.5){\ssp}\put(1,-.5){\ssm}
    \put(0.6,0.25){\line(1,0){.8}}\put(0.6,-.25){\line(1,0){.8}}}
    \end{picture}
    \simeq\; {\cal G}^{--}_{res}{\cal A}^{-+}{\cal G}^{++}_{res}\simeq
    \begin{picture}(1.4,1)\put(-0.3,.15){\put(.6,.75){\line(1,0){.8}}
    \put(.6,-.75){\line(1,0){.8}}
    \put(1,0){\interaction}}\end{picture}
\end{equation}
with like--sign effective interactions (\ref{4point}) (whereas
in the general case one comes to  (\ref{open}) with renormalized
${\cal A}^{-+}_{ren}$ loops).

\section{Contribution of More Complicated Diagrams}
Neglecting vertex corrections
diagram (\ref{corr2}) is given as
\begin{eqnarray}\label{ImA1}
  \unitlength6mm\begin{picture}(3.5,1.)
  \put(0,0.15){\oneloopvertex}\put(1.625,0.15){\interaction}
  \end{picture}&=&-\ii\Pi^{-+}_{n=1}(q)
  =-\ii V^{\mu}V^{\nu}{\cal A}^{- + }_1(q)
\; ,\where\nonumber\\[5mm]
-\ii{\cal A}^{- +}_1(q)&=&-\int \frac{d\epsilon d^3p}{(2\pi)^4} \frac{d\omega
d^3k}{(2\pi)^4}\ii G^{- -}(p+k)\ii G^{+ +}(p-q) \\
&&\times \ii G^{- +}(p)\ii G^{+ -}(p+k-q)
\ii {\cal G}^{- +}(k)\nonumber\;,
\end{eqnarray}
with four-vectors $q=(\omega_q,{\vec q})$, $p=(\epsilon,{\vec k})$ and
$k=(\omega,{\vec k})$.
Here ${\cal G}^{- +}$ is the "-- +" interaction loop (\ref{open}).
For $q\ll k\sim p_F$ the integration in (\ref{ImA1}) over the
$\vec{p}\vec{k}$--angle gives\footnote{using also $q$, $k$ and $p$ for
$|{\vec q}|$, etc.}
\begin{eqnarray}
\label{J0}
J_0(\epsilon)&=&
\int_{-1}^{1} dx G^{- -}(p+k)G^{+ -}(p+k-q)\nonumber \\
&=&
(1-n_{\epsilon+\omega-\omega_q})\frac{2m_F^{\ast}\pi}{pk}
\frac{-\omega_q+\ii\Gamma\; \mbox{th}((\epsilon+\omega)/(2T))}
{\omega_q^2+\Gamma^2}.
\end{eqnarray}
Since $\mid x\mid <1$ one has the following restrictions on
$p$
\begin{equation}\label{epsilon0}
\epsilon_p=p^2/2m_F^{\ast}>\epsilon_0=(\omega-k^2/2m_F^{\ast}-
\omega_q)^2m_F^{\ast}/2k^2.
\end{equation}
Further integrations can only be done in certain limits.  For $\Gamma,
\omega_q \ll T$ one can use Kadanoff--Baym ansatz (\ref{ap3}),
c.f. Appendix A, (\ref{KBa}). In that case the $\epsilon$ integration
can be performed.  One needs only the real part of expression
(\ref{ImA1}) since the imaginary part is cancelled by the
corresponding diagram with opposite time ordering (opposite line
sense). Thus,
\begin{eqnarray}
\label{J1}
J_1&=&\Re
\int_{-\infty}^{\infty} d\epsilon G^{+ +}(p-q)G^{- +}(p)
J_0(\epsilon)
\cr
&\simeq&
n_{\epsilon_p-\mu_F}
(1-n_{\epsilon_p-\mu_F+\omega-\omega_q})
\frac{2m_F^{\ast}\pi^2}{pk}
\frac{\omega_q^2-\Gamma^2 u}
{(\omega_q^2+\Gamma^2)^2},
\end{eqnarray}
where
\begin{eqnarray}\label{th}
u&=&\mbox{th}\frac{\epsilon+\omega}{2T}
\mbox{th}\frac{\epsilon-\omega_q}{2T}=
1-2n_{\epsilon-\omega_q}
(1-n_{\epsilon+\omega})-2n_{\epsilon+\omega}(1-n_{\epsilon-
\omega_q})
\\ &\simeq& 1-2n_{\epsilon_p-\mu_F-\omega_q}
(1-n_{\epsilon_p-\mu_F+\omega})-2n_{\epsilon_p-\mu_F+\omega}
(1-n_{\epsilon_p-\mu_F-\omega_q}).\nonumber
\end{eqnarray}
For low fermion occupations $u$ is about unity and
we obtain with the help of  relation (\ref{nrel})
\begin{eqnarray}
\label{A1res}
-\ii{\cal A}^{- +
}_1&=&C_1(\omega_q)\int \frac{m_F^{\ast 2}}{16\pi^4\omega_q^2}
n^B_\omega  n^B_{\omega_q-\omega}\Im \Sigma_{0,B}^{R}(\omega,k)
J_2(\omega,k)kdk d\omega,\quad\mbox{with}\\
J_2(\omega,k)&=&\int_{\epsilon_0}^{\infty}
(n_{\epsilon_p-\mu_F+\omega-\omega_q}-n_{\epsilon_p-\mu_F})
d\epsilon_p,\quad
  C_1(\omega_q)=\omega_q^2
\frac{\omega_q^2-\Gamma^2}{(\omega_q^2+\Gamma^2)^2}.\nonumber
\end{eqnarray}
The integral in eq. (\ref{A1res}) can be expressed through
${\cal A}_0^{-+}$ in QPA (eq. (\ref{ImAT})), since $pk\gg{m^{\ast}}\Gamma$
and one obtains
\begin{eqnarray}\label{A1tot}
-\ii{\cal A}^{- +}_1(\omega_q,{\vec q})=C_1(\omega_q)
&\int &\frac{{\cal G}^{+ +}(\omega,k){\cal G}^{- -}(\omega,k)}{4\pi^3}
n^B_\omega  n^B_{\omega_q-\omega}\\ &&\times\Im {\cal A}_0^{R}(\omega,k)
\Im {\cal A}_0^{R}(\omega_q-\omega,k)k^2 dk d\omega.\nonumber
\end{eqnarray}
This expression differs from the contribution of the corresponding QPA
Feynman diagram calculated in ref. \cite{V3} only by the pre--factor
$C_1(\omega_q)$ which is non--unit in our case of finite width
$\Gamma$.  In the QC limit this expression (\ref{A1tot}) coincides
with the $n=1$ term in classical Langevin result.

Diagram
\begin{equation}
  \unitlength6mm
  \begin{picture}(3.8,1.)
  \put(0,0.15){\doubleloop}\end{picture}=-\ii{\cal A}^{-+}_{\mbox{ex}}
\end{equation}
can be evaluated along similar lines considering the integrals for the
left and right sub-loops
\begin{eqnarray}
\label{loop1}
J_3
&=&\int (-1)iG^{- -}(p+k)iG^{- +}(p)iG^{+ -}(p+k-q)\frac{d^4p},
{(2\pi)^4}\\
\label{loop2}
J_4
&=&\int (-1)iG^{+ +}(p_1+k-q)iG^{- +}(p_1+k)iG^{+ -}(p_1)
\frac{d^4p_1}
{(2\pi)^4}.
\end{eqnarray}
Integration of (\ref{loop1}), (\ref{loop2}) is done quite analogously
to the previous cases. One may integrate over $\vec{p}\vec{k}$--angle,
then over $\epsilon$ and $\epsilon_p$ in eq. (\ref{loop1}) and over
$\vec{p_1}\vec{k}$--angle, $\epsilon_1$ and $\epsilon_{p1}$ in
eq. (\ref{loop2}), respectively. One recovers the corresponding QPA
form derived in ref. \cite{V3}, however multiplied by the pre--factor
$C_0(\omega_q)$, c.f. (\ref{corr2}).

\noindent {\bf Acknowledgments:} We thank G. Bertsch, J. Bondorf,
J. Cleymans, P. Danielewicz, B. Friman, K. Geiger, M. Gyulassy,
P. Henning, M. Herrmann, Yu. Ivanov, J. Kapusta, I. Mishustin,
B. M\"uller, V. Toneev and X. N. Wang for valuable suggestions and
remarks at various stages of this study.  D. N. V. thanks GSI for
hospitality and support.  The research described in this publication
was also made possible for him in part by Grant N3W000 from the
International Science Foundation and the Russian Government.\\[-12mm]

\end{document}